# Energy-efficient, Reconfigurable Optoelectronic Artificial Synapses Based on $MoWS_2$ Alloy for Pattern Recognition and Color Image Filtering Applications

Deepak Kumar Sahu[1], Santu Kumar Ghosh[1], Sagarneel Ghoshal[1], Saranya Das[1], Samit K. Ray[1,*]

[1]*Department of Physics, Indian Institute of Technology Kharagpur, Kharagpur, India, 721302*

[*]*Corresponding author email: physkr@phy.iitkgp.ac.in*

## Abstract:

Two-dimensional transition-metal dichalcogenide alloys are potential candidates for advanced optoelectronic and neuromorphic applications due to their strong light-matter interactions and controllable defect properties. However, large-area growth of such alloys remains challenging, while the correlation between their physical and neuromorphic properties remains largely unclear. In this work, we present an innovative microcavity chemical vapor deposition (CVD) reactor pathway to grow uniform, and large-area $MoWS_2$ mono- and few-layer alloy films for demonstrating optoelectronic synaptic functionalities. Driven by growth-induced intrinsic sulfur vacancies, as confirmed by XPS, KPFM, and STEM measurements, our optoelectronic synaptic device (OSD) successfully emulates essential biological synaptic features, such as excitatory postsynaptic currents (EPSC), paired-pulse facilitation (PPF~170%), and stimulus-dependent short- and long-term plasticities (STP & LTP). With picojoule-order energy consumption per synaptic event and nanoampere-order dark current, the device enables low-power neuromorphic learning, including emulation of Pavlovian associative learning. Furthermore, the experimentally measured conductance weight-update characteristics enabled an artificial neural network (ANN) simulation to achieve 92.43% recognition accuracy on the MNIST handwritten digit dataset. Finally, we demonstrate advanced neuromorphic visual processing by executing color image filtering based on the device's wavelength-selective photoresponse characteristics. This simple, yet multifunctional device architecture provides a promising path toward energy-efficient, spectral-selective neuromorphic vision applications.

## Introduction:

To overcome the memory bottleneck and process latency of conventional von Neumann computing, energy-efficient neuromorphic computing systems are being studied extensively, which enable parallel, intensive data processing and adaptive learning capabilities, providing hardware support by mimicking biological synapses that integrate signal sensing, computation, and memory into a single entity[1–4]. Traditional von Neumann computing architecture faces critical challenges in data transfer, such as increased process latency and energy consumption, owing to the physical separation of processing and memory units. By integrating memory and processing into a single unit, it dramatically reduces the energy consumption per synaptic event. The human nervous system possesses a natural neural network (NNN) comprising more than ~$10^{11}$ neurons and ~$10^{15}$ synapses[5]. Synapses are specialized junctions between neurons that play a vital role in neural functioning by facilitating the transmission of neural signals using neurotransmitters, from the pre-synapse to the post-synapse in response to external stimulation, thereby triggering the next neuron to fire. These synapses can dynamically adjust their responsiveness to external stimuli and training, or brain activity. The change in synaptic weight (connection strength between two neurons, either strengthening or weakening) is called synaptic plasticity, and it depends on the synapse's activity history, playing a key role in cognitive functions and behaviour. An analogy can be made between a biological synapse and an artificial synapse, particularly in terms of changes in specific parameters, such as conductance, which can be treated as excitatory postsynaptic current (EPSC) or inhibitory postsynaptic current (IPSC), in response to external perturbations. Among human brain-mimicked neuromorphic systems, the optoelectronic synaptic device (OSD) is considered a new paradigm due to its ultrafast signal transmission, large bandwidth, low crosstalk, low power consumption, and wireless communication capabilities, utilizing light as an input stimulus instead of electrical pulses. In an artificial OSD device, measuring photocurrent upon optical excitation is a practical approach to replicating biological synaptic functionalities.

Recently, two-dimensional (2D) transition metal dichalcogenides (TMDs) have attracted significant research attention for ANN applications, such as pattern recognition and neuromorphic vision systems, due to their exciting electronic and optical properties, such as strong light-matter interactions, tunable band gaps, extraordinary stability, and controllable defect characteristics[6]. Alloying offers a robust way to intrinsically control the structural and electronic properties of 2D materials. Alloying of 2D binary TMDs can further modulate the bandgap and facilitate defect engineering by varying the composition of either the chalcogen or the transition-metal component, and by introducing controlled lattice disorder due to differences in their atomic radii and lattice parameters. Without the support of external vacancy generation processes, such as ion bombardment or gas adsorption, the intrinsic lattice distortion and growth-induced sulfur vacancies in the resultant ternary alloy may act as charge trapping/de-trapping centres to realize the optoelectronic synaptic memory effect upon light exposure. Therefore, developing a synaptic device based on 2D ternary alloys that offers multiwavelength response with a facile device structure and low power consumption is essential for the advancement of optical neuromorphic systems. Moreover, TMD alloys exhibit a composition-dependent bandgap gradient, facilitating the separation of photo-generated charge carriers, thereby prolonging their lifetimes. Chemical vapor deposition (CVD) technique facilitates the alloying of TMDs due to the abundance of intrinsic sulfur vacancies, with a lower driving force than in mechanically exfoliated flakes[7].

In this study, a specialized microcavity reactor-assisted growth strategy has been employed to synthesize uniform monolayer/few-layer $Mo_{0.5}W_{0.5}S_2$ ($MoWS_2$ ) alloy films. The microcavity-assisted space-confined growth overcomes certain challenges, such as non-uniformity, limited growth size, and reproducibility issues in the synthesis of TMD alloys, which are highly sensitive to growth conditions. Under optimized conditions, a large-area, few-layer alloy $MoWS_2$ film on $SiO_2$/Si has been developed to fabricate gate-free, two-terminal optoelectronic synaptic devices. Notably, the device structure is quite simple, but with $MoWS_2$ as the only active light-sensitive material, it exhibits optical synaptic behaviour, mimicking all the major functionalities of the human nervous system, such as EPSC, PPF, STP & LTP, short-term and long-term memory (STM & LTM) across various optical pulse specifications, such as pulse width, pulse number, optical stimulation frequency, optical power, electrical read bias, etc. In addition, typical learning-forgetting behaviour mimicking human brain's perpetual learning process, including associative learning via Pavlov's dog experiment is demonstrated. With a minimum power consumption of 239 pJ per single optical spiking event, the synaptic device enables a low-power-consuming neuromorphic computing architecture. An ANN simulation has been carried out on the Modified National Institute of Standards and Technology (MNIST) handwritten dataset for the digit recognition task, and recognition accuracy of ~93% has been obtained using the experimentally measured conductance-weight update data. By leveraging synaptic plasticities at different wavelengths in our device, we have demonstrated color image filtering (feature extraction from a mixed-color image), thereby making our single-material platform attractive for emerging low-power, high-speed neuromorphic computing and color-vision applications.

# Results and Discussions:

## a) Chemical vapor deposition growth of $MoWS_2$ alloy:

As shown in Figures 1a-c, the growth of alloyed $MoWS_2$ was carried out in a rectangular space-confined microreactor CVD chamber[8,9]. Before heating, the quartz tube was thoroughly cleaned and pumped down to remove moisture, oxygen, and other gaseous contaminants. The space-confined microreactor chamber was meticulously designed to enable uniform growth of the alloy $MoWS_2$ over a large area, in contrast to the conventional direct growth approach (with the front side facing the reacting precursors), to compare the advantages and limitations of both configurations under identical growth conditions. Growth temperature profiles of both S and $WO_3$/$MoO_3$ in the modified space-confined micro-reactor CVD are shown in Figure S1a. The details of the CVD growth mechanism are described in Supplementary Note 1. We observed that the space-confined microcavity approach yields highly uniform, large-area monolayer/few-layer coverage (Figure

S1b). In contrast, the conventional approach ($SiO_2$/Si facing upside down; obtained by simply removing the bottom $SiO_2$/Si substrate from the microcavity stack) produced a few-layer-thick non-uniform flakes across the substrate surface (Figure S1c). Under suitable conditions, both $WO_3$ and $MoO_3$ vapors overlap, leading to large-area film growth of $MoWS_2$ alloy. On the other hand, the microcavity configuration confines incoming precursor vapors, resulting in a uniform monolayer film of the alloy. Inside the cavity reactor chamber, Mo and W-containing vapors accumulate efficiently. That promotes vertical growth after initial monolayer formation and second-layer nucleation, thereby enabling controlled few-layer stacking. Microcavities also help maintain the uniform sulfur chemical potential locally and form a quasi-equilibrium local environment. But in the direct growth approach via APCVD, or without a microcavity reactor, the S vapors fluctuate locally on the substrate surface, leading to uncontrolled sulfurization and non-uniform growth. To grow large-area uniform thin films instead of individual grains, we systematically optimized several key growth parameters. The formation of $MoWS_2$ flakes or their coalescence in the form of a thin film is a result of both thermodynamic and kinetic equilibrium of the growth process. Hence, we optimized key growth metrics, such as growth temperature and gas flow rate (Figure S2), to regulate the above equilibrium states. $MoS_2$ and $WS_2$ were also synthesized as control samples through a similar process. An optical micrograph of $MoWS_2$ flakes of different thickness and a topographic atomic force microscopy (AFM) image of the monolayer flake (thickness ~0.8 nm) are shown in Figure 1d and e, respectively.  Figure 1f presents a comparative plot of the Raman spectra for binary $MoS_2$, $WS_2$, a heterostructure of $MoS_2$-$WS_2$ and ternary $MoWS_2$ alloy. The emergence of three distinct, characteristic Raman modes (Figure 1f, top panel) confirms the successful formation of the ternary alloy in contrast to the standard two-mode features of parent materials ($MoS_2$ and $WS_2$). In the ternary alloy, $WS_2$ and $MoS_2$-like $E^1_{2g}$ modes appear at ~350 $cm^{-1}$ and ~380 $cm^{-1,}$ respectively, while the combined $A_{1g}$ mode for both $MoS_2$ and $WS_2$ appears at ~405 $cm^{-1}$. As shown in Figure 1f, pure $MoS_2$ exhibits two vibrational Raman modes at 383 $cm^{-1}$ (in-plane vibration) and 405 $cm^{-1}$(out-of-plane vibration), while the same for $WS_2$ appears at 350 $cm^{-1}$ and 418 $cm^{-1}$, respectively. On the other hand, four Raman modes consisting of two each from parental material are observed in $MoS_2$-$WS_2$ heterostructure. The Raman results clearly display three distinguishable characteristic peaks, revealing the formation of a ternary $MoWS_2$ layer. The Raman intensity map of the $A_{1g}$ mode of a monolayer $MoWS_2$ is shown in Figure 1g, indicating uniform film growth. The photoluminescence spectra for monolayer $MoS_2$, $WS_2$, and ternary $MoWS_2$ alloy are displayed in Figure 1i. Other Raman modes ($MoS_2$ and $WS_2$-like $E^1_{2g}$ modes) and PL mapping of the A excitonic peak for the monolayer $MoWS_2$ triangular flake, presented in Figure S5, strongly support uniform alloy growth. The optical bandgap of the alloyed $MoWS_2$ (1.85 eV) is found to lie between that of $MoS_2$ (1.82 eV) and $WS_2$ (1.88 eV), consistent with Vegard's law[10]. The EDS elemental mapping of Mo, W and S for $MoWS_2$ thin film is shown in Figure S3 (d-g).

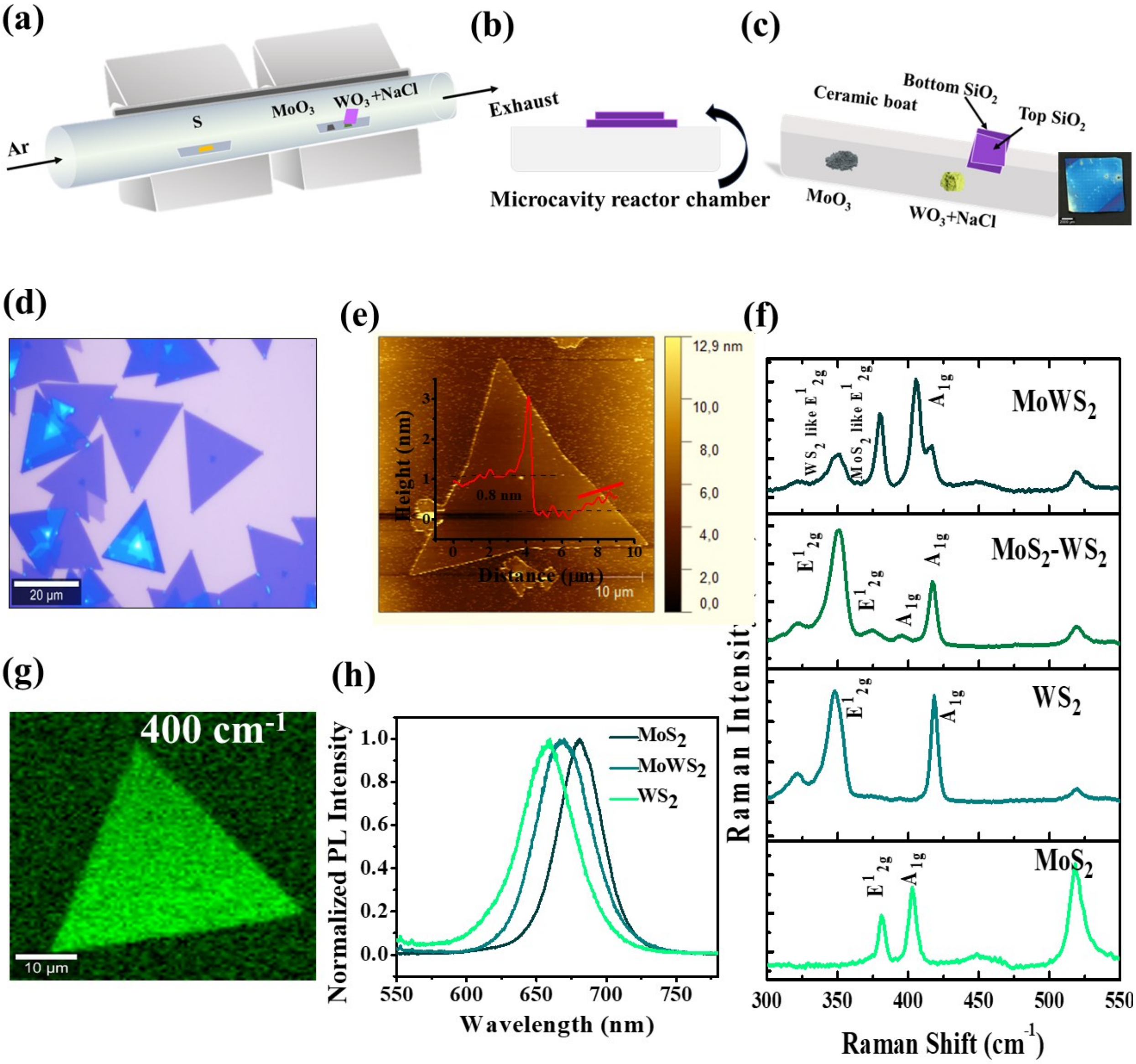

*Figure 1: (a) Chemical vapor deposition set up used for the growth of $MoWS_2$ alloy under low-pressure condition, (**b**)Configuration of microcavity reactor inside the growth chamber, (c) Schematic location of the precursors inside the growth chamber with CVD-grown uniform film in the inset, (d) Optical image of triangular $MoWS_2$ flakes of varying thickness on 300 nm $SiO_2$ /Si substrate (e) AFM topographic image of a triangular monolayer $MoWS_2$ flake, with thickness ~0.8 nm (f) Comparative Raman spectrum of ternary $MoWS_2$ alloy along with $MoS_2$, $WS_2$ and their $MoS_2$-$WS_2$ heterostructure (g) Raman mapping of $A_{1g}$ mode of $MoWS_2$ alloy shown in Figure 1(f). (h) Photoluminescence spectra of ternary $MoWS_2$, relative to parental compositions $MoS_2$ and $WS_2$.*

Thin crystalline quality and phase purity of CVD-grown $MoS_2$, $WS_2$ and $MoWS_2$ films have been studied using a high-energy synchrotron X-ray diffraction (XRD) system with X-rays of wavelength ~1.01 Å and fluence ~ $10^{10}$ photons/seconds (instead of standard Cu Kα source X-rays, λ ~1.5406 Å). Grazing-angle (0.2º) beamline synchrotron XRD spectra of CVD-grown films are shown in Figure 2a, with the characteristic (002) peak, indicating a layered hexagonal structure with ordered c-axis orientation, appearing for all three samples. The ternary $MoWS_2$ layer exhibits a single, but quite broadened, (002) peak at 2θ~ 9.45º, compared to MoS2 (2θ~ 9.63º) and WS2 (2θ~ 9.59º). A small peak shift is attributed to the alloying of two different-sized transition-metal atoms, Mo and W, via atomic substitution. After Rietveld refinement, the full-width half- maximum (FWHM, $\beta$) values obtained for the characteristic (002) peak for $MoS_2$, $WS_2$ and $MoWS_2$ are 0.6024º, 0.1036º and 1.0362º, respectively. A higher FWHM value for the ternary TMD, compared to its binary counterparts, is ascribed to the presence of finite but random disorder in the alloy system. The absence of any other diffraction peaks in the XRD spectra indicates the growth of a thin film with highly preferred orientations solely in the 2H-phase of the TMD alloy. A second-order, weak (004) peak for the 2H phase is observed at 2θ ~18° and corresponds to the higher-order reflection of the (002) plane. The presence of the higher-order reflection plane corresponding to the (004) peak is attributed to ordered c-axis growth and improved crystallinity, but it is less prominent due to the lower film thickness.

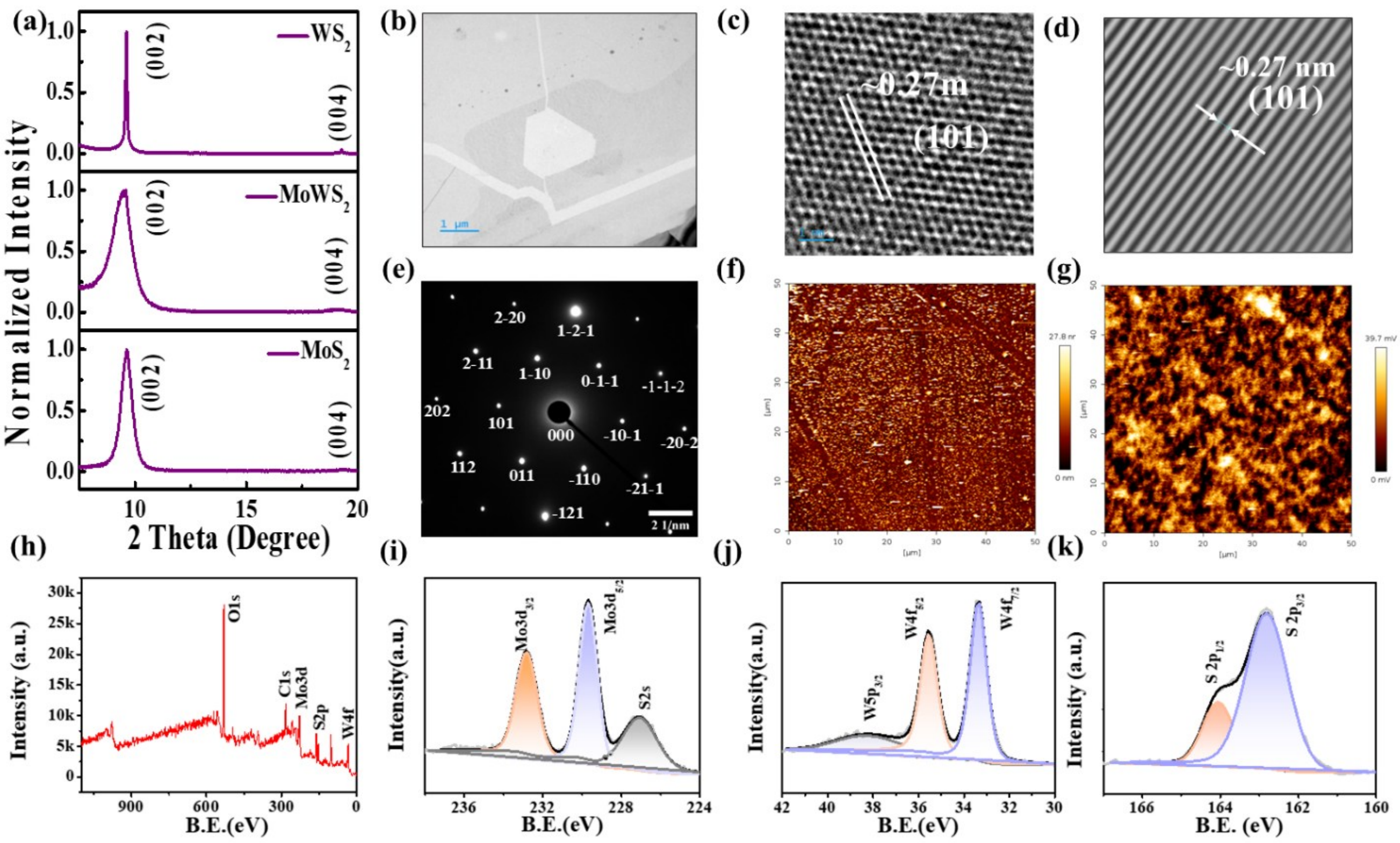

*Figure 2: (a) Synchrotron XRD spectra of $MoS_2$, $WS_2$, and $MoWS_2$ alloy films. (b) Low-resolution TEM image of the transferred $MoWS_2$ flake on $SiO_2$/Si substrate. (c) High-resolution TEM image of the alloyed $MoWS_2$ flake, with visible lattice fringe pattern, depicting an interplanar spacing of 0.27 nm, corresponding to the (101) crystal plane. (d) The Inverse FFT pattern, generated from the TEM image shown in Figure 2c,*

*exhibiting distinct fringes. (e) SAED image with a distinct hexagonal pattern, depicting growth of single-crystalline few-layered $MoWS_2$ (f) Typical AFM surface topographic image of grown $MoWS_2$ film (g)Surface potential mapping of the same region of Figure (f) by KPFM measurement (h) XPS survey scan of the $MoWS_2$ alloy and (i, j and k) high-resolution XPS spectra for Mo, W and S elements showing the binding energy for corresponding electrons of the 2H-phase ternary alloy*

There is a natural tendency for sulfur vacancies to form during CVD growth of the $MoWS_2$ alloy due to differences in lattice parameters between $MoS_2$ and $WS_2$, thereby avoiding the need for complex ex-situ techniques such as high-energy ion bombardment. A high-resolution transmission electron microscope (HRTEM) was used to study the nanoscale structural properties, including crystal structure, lattice fringes, stacking order, grain boundaries, and defect centres. Also, sulfur vacancies, among other defect types, have the lowest formation energy and are readily formed during alloying at higher growth temperatures, as confirmed by HRTEM and STEM imaging. Figure 2b shows the low-resolution TEM image of the $MoWS_2$ flakes transferred on a copper grid using the surface-energy-assisted wet-transfer technique [11]. Figure 2c shows the high-resolution TEM image of the flake with distinctly visible sulfur vacancies, which is clearly shown in the aberration-corrected HAADF image (Figure S4a), with brighter spheres in the hexagonal unit cell representing the transition elements (more bright means higher atomic number element Mo or W), while the less bright ones represent the sulfur atoms. The interplanar spacing between the lattice fringes is found to be ~ 0.27 nm, corresponding to the (101) crystal plane of the 2H-phase $MoWS_2$ alloy, which is consistent with the Inverse Fast Fourier Transform (IFFT) results presented in Figure 2d. The high-resolution TEM micrograph provides preliminary evidence of the existence of structural defects. Sulfur vacancies are the most common and inevitable defects in TMDs due to their low formation energy, alloying with two different-sized atoms, non-equilibrium growth conditions, and substrate-lattice mismatch[7,12–16]. Furthermore, the selected-area electron diffraction (SAED) pattern (Figure 2e) of the $MoWS_2$ film reveals the formation of the 2H-phase with high crystallinity and preferred orientation, as evidenced by the presence of hexagonal dot patterns. The diffraction pattern could be indexed as (101), (011), (-110), (-10 1), (0-1-1), and (1-10) Miller indices, while the central region, labelled as (000), was blocked by the shutter. The crystal plane indexing of the SAED pattern has been performed using the Crystbox software along the (001) zone axis, as shown in Figure 2e.

Kelvin Probe Force Microscopy (KPFM) imaging was conducted to understand the nature of the trap state that captures free carriers via surface potential analysis. Figure 2f shows the AFM topographic image of the $MoWS_2$ film surface. The height profile analysis confirms that the sample is topographically uniform, whereas Figure 2g displays the surface potential map of the same region as shown in Figure 2f. The surface potential map shows a completely different picture of the film, with a highly non-uniform distribution of surface potential. The presence of several small bright and dark regions in the surface potential map (Figure 2g) indicates a non-uniform distribution of the electron and hole trap states. The role of these trap centres in the synaptic activity of $MoWS_2$ film is explained later in this manuscript.

The existence of sulfur vacancies, acting as defect centres in the $MoWS_2$ lattice, can also be corroborated from the Raman spectrum (Figure 1f) by monitoring the shift of the dominant Raman modes of $MoWS_2$ ($E^1_{2g}$ and $A_{1g}$-like modes for the ternary alloy), as detailed below: (i)the full-width-half-maximum (FWHM) is increased for both modes, (ii) $E^1_{2g}$ mode exhibits a redshift caused by the phonon confinement[17,18]. Moreover, the presence of longitudinal acoustic mode 2LA (M) mode at ~450 $cm^{-1}$ (Figure S5) can also be a strong indication of defect formation in the $MoWS_2$ lattice, as reported elsewhere [19,20]. The broadening of the prominent 2LA(M) peak corresponds to the vacancy formation, especially the sulfur vacancy, due to their low formation energy, which changes the bond length of the Mo-Mo bond as well as the Mo-S bond. Our observation is in strong agreement with previous reports showing a contraction of Mo-Mo (1%) and Mo-S2S (4%) bond length, caused by the presence of sulfur vacancies[21,22].

Furthermore, the elemental analysis and chemical oxidation states of the as-grown $MoWS_2$ sample have been investigated using X-ray photoelectron spectroscopy (XPS). Figure 2h displays the survey spectrum for $MoWS_2$ alloy, confirming the presence of Mo, W, and S elements without any impurities except the surface adsorbed C and O. Figure 2 i, j and k show the high-resolution XPS spectra displaying the binding energy of Mo 3d, W 4f and S 2p electrons in the directly grown $MoWS_2$ layer. All XPS peak fittings have been performed using XPS PeakFit software, with Shirley background and carbon reference corrections. The characteristic binding energy values for Mo 3d are 232.9 eV and 229.8 eV, corresponding to the $Mo3d_{3/2}$ and $Mo3d_{5/2}$ states, respectively, which strongly corroborate the existence of the Mo $4^{+}$ state. It also includes the S1s peak at 227.1 eV. Similarly, the binding energy values for the deconvoluted W 4f peaks located at 35.6 eV and 33.3 eV, corresponding to $W4f_{5/2}$ and $W4f_{7/2}$, respectively, confirm the presence of W in $4^{+}$ state. The pristine $WS_2$ has been found to possess two chemical states (32.6 eV for $W4f_{7/2}$ and 34.7 eV for $W4f_{5/2}$), as reported previously[23]. The shift in the binding energies of both Mo3d and W4f electrons is attributed to the formation of a ternary 2D alloy with modified chemical bonding. On the other hand, the S2p spectrum exhibits well-resolved spin-orbit splitting induced doublet peaks at 164.1 eV and 162.8 eV, corresponding to the $S2p_{3/2}$ and $S2p_{1/2}$ electrons, respectively. These peaks are characteristic of the S(II) or $2^{-}$ valence states[24,25]. The stoichiometry of the $Mo_xW_{1-x}S_2$ alloy from XPS spectra has been estimated to be 0.43:0.47:1.78, from the spectral intensities of all contributing elements, which clearly demonstrates sulfur deficiency in the CVD grown sample (See Supplementary Note 2). It is known that sulfur vacancies in $MoWS_2$ introduce localized trap states within the bandgap, acting as effective charge-trapping centres under optical excitation. Upon illumination, the photogenerated charge carriers can be captured by these vacancy-induced mid-gap states, leading to persistent photoconductivity and a gradual modulation of channel conductance. Thus, sulfur vacancies in CVD-grown 2D ternary alloys may play a crucial role in enabling energy-efficient, light-driven synaptic functionality in optoelectronic devices.

## b) Light response characteristics of the optical synaptic devices:

Figure 3a illustrates the schematic of a biological synapse and the process of signal transmission that occurs across it. The analogous $MoWS_2$-based artificial OSD schematic, shown as a subfigure in Figure 3a, essentially represents the schematic of the as-fabricated two-terminal device with a metal-semiconductor-metal (MSM) type configuration, with a very thin layer of $MoWS_2$ (thickness ~6 nm and roughness of the film ~1 nm, along with FESEM image of the thin film, Figure S6) directly grown on $SiO_2$ (300 nm)/Si substrate as the principal functional light-absorbing layer, along with two Cr/Au electrodes. An optical pulse acts as a presynaptic spike, enabling the generation of photoinduced carriers and their migration across sulfur vacancies within the $MoWS_2$ layer. This process alters the channel conductance (here, resembling synaptic weight) and ultimately generates an EPSC (post-synaptic spike) within the active channel (length: 100 µm), which slowly decays upon removal of the optical source. Figure S3 (a-c) shows optical microscopy images of the fabricated OSD with two Cr/Au electrodes.

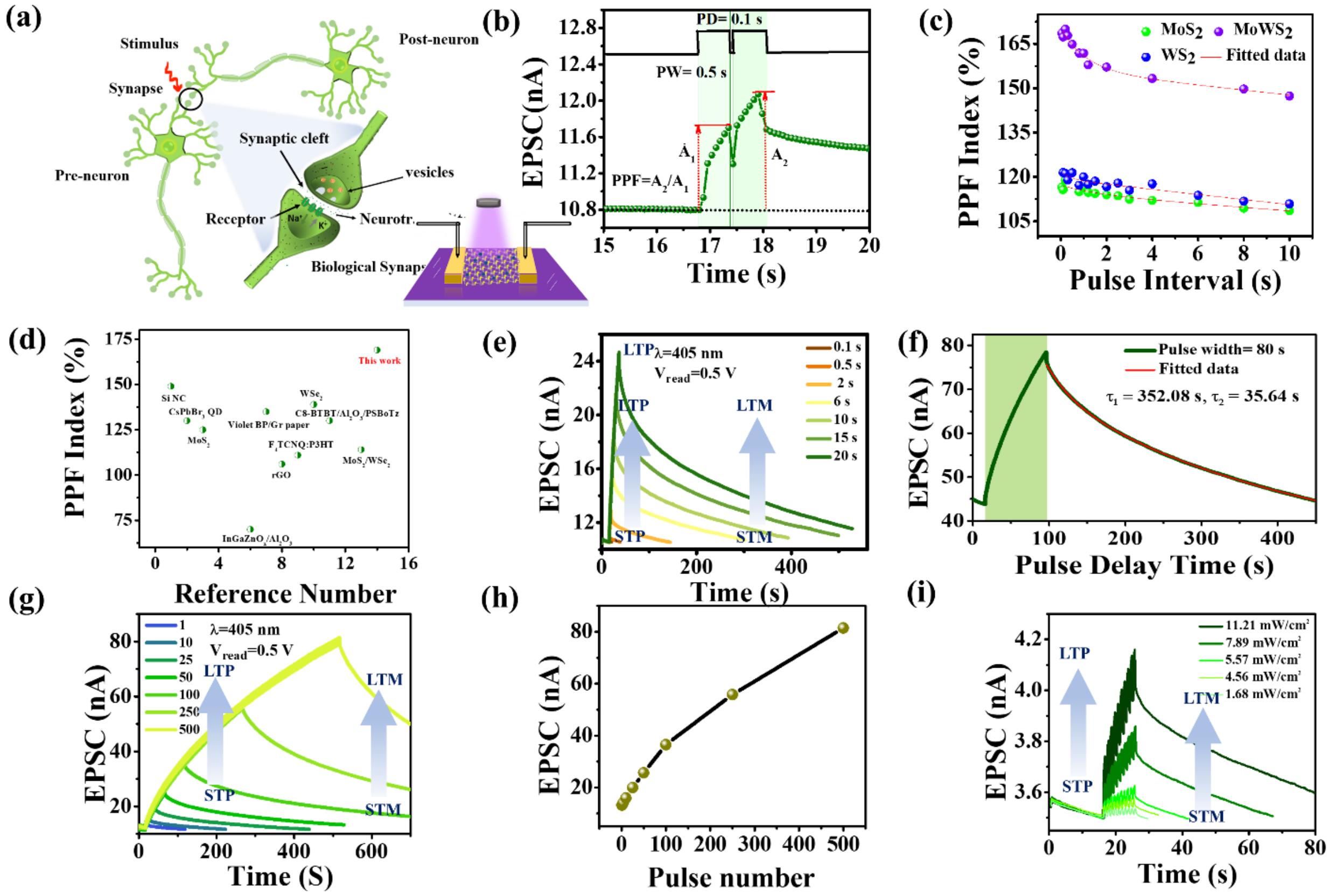

*Figure 3: (a) Schematic of biological synapse with neurons and synapses; the inset shows the analogue artificial synapse with $MoWS_2$ layer (b) The paired pulse facilitation with two consecutive pulses (width= 500 ms, interval 100 ms) (c) Variation of PPF indexon the delay time for $MoS_2$, $WS_2$ and $MoWS_2$ layers (d)Comparison of our device performance in terms of PPF index with reported results; $MoS_2$[26], Si NC[27], $CsPbBr_3$ QD[28], violet BP[29], $WSe_2$[30], C8-BTBT/$Al_2O_3$/PSBoTz[31], FTCNQ:P3HT[32], rGO[33], $MoS_2$/$WSe_2$[34], $InGaZnO_x$/$Al_2O_3$[35] (e ) STP-to-LTP transition with increasing pulse duration ( spike duration dependent plasticity) (f) Long-term potentiation and the memory retention after removal of optical signal (g) STP-to-LTP transition with increasing pulse number (spike number dependent plasticity) (h) Variation of EPSC with pulse number (i) STP-to-LTP transition with increasing optical power density ( a train of 10 pulses; each with PD and PW= 500 ms)*

To investigate the effect of alloying in TMDs, we also fabricated two control devices with binary $MoS_2$ and $WS_2$ as the channel material and compared their key synaptic features. From the current-voltage characteristics (Figure S8a), the lowest dark current observed in the alloy device compared to the binary TMD-based OSDs is attributed to the suppression of deep-level defect states (DLDS) via alloy engineering, thereby reducing the non-radiative recombination rate. On the other hand, deep-level defect states that act at Shockley-Reed-Hall recombination centres, resulting in higher dark current [36,37]. This result is significant because it enables the ternary device to operate with low power consumption, providing a powerful solution to the energy challenge that modern neuromorphic computing technology currently faces. As mentioned earlier, synaptic plasticity is an important characteristic of OSDs and can be classified into short-term plasticity (STP) and long-term plasticity (LTP) based on the duration of optical memory retention[38]. STP refers to the temporary change in the synaptic weight that can be sustained for a short span of time (a few milliseconds to a few minutes), responding to a series of closely spaced short optical pulses and plays a crucial role in information processing and memory[39,40]. Paired-pulse facilitation (PPF) phenomenon is a distinctive example of an STP event in the human brain. Whereas LTP corresponds to the change in synaptic strength that endures for a few minutes to hours or even a lifetime, which is the basis for learning and memory in neuromorphic systems[41]. Figure 3b shows the PPF characteristics subjected to two successive short light pulses (pulse width 500 ms, 405 nm wavelength, power ~10 mW/cm$^2$) with a 100 ms pulse delay. It is found that the response to the second stimulus is significantly higher than that to the first. PPF is a form of short-term plasticity that demonstrates how synapses conduct signal transmission when short-pulse stimuli are consecutively applied. The PPF index is defined as the ratio of the response to the second stimulus to the response to the first stimulus [42,43]

$$PPF = \frac{A_2}{A_1} \times 100\% \qquad \text{.........[1]}$$

, where $A_1$ and $A_2$ represent EPSC amplitudes for first and second optical spikes, after subtracting the initial dark current. The PPF effect in our device, shown in Figure 3b, is due to the electron-trapping phenomenon, where certain photogenerated charge carriers are captured after the initial pulse, resulting in a metastable current. Thus, the second pulse produces additional charge carriers, resulting in a second EPSC spike that is greater than the first[44]. The PPF index values for different time intervals (Δt) have been evaluated and plotted as a function of Δt, as depicted in Figure 3c. Since Δt is evidently less than the EPSC decay time, the PPF index is quite sensitive to Δt. However, the PPF index significantly diminishes with an increase in Δt, resulting in a fast decay process characterized by a relaxation time $\tau_1$. When Δt exceeds the EPSC decay time, the amplification of the photocurrent induced by the second spike completely diminishes. In our devices, we observed a maximum PPF index of 170% over a time interval (Δt) of 100 ms between two consecutive pulses. In comparison, the PPF index drastically drops to 121% and 118% for control $MoS_2$ and $WS_2$, respectively, over the same time interval (Δt = 100 ms) between two successive pulses, as shown in Figure 3c. Thus, alloyed $MoWS_2$ exhibits better PPF performance than its parental compounds. However, as the time interval increased from 100 ms to 8 s, the PPF index of the alloy device decreased from ~170% to ~150% (Figure 3c and S9a). As presented in Figure 3d, our ternary $MoWS_2$ device exhibits the highest PPF index value of ~170% for a pulse delay of 100 ms in comparison to other devices reported to date[45]. The PPF performance of our device has been compared with that of existing devices in the literature and is shown in Figure 3d. In our ternary 2D alloy device, the decrease in the PPF index with Δt may be attributed to electron trapping and disappearance over the longer time interval. The decay of the PPF with Δt can be well-fitted by a double-exponential decay function[43], as follows.

$$PPF = 1 + C_1 \exp\left(\frac{-\Delta t}{\tau_1}\right) + C_2 \exp\left(\frac{-\Delta t}{\tau_2}\right) \dots\dots\dots[2]$$

Where $C_1$ and $C_2$ are fast and slow facilitation constants, and $\tau_1$ and $\tau_2$ are the corresponding characteristic relaxation time constants for the fast and slow decay phases. After fitting the PPF decay curve for $MoWS_2$ sample, we obtained the relaxation time constant values as $\tau_1$ =1.188 s and $\tau_2$=223.36 s.

Few-layer $MoWS_2$ with an optical bandgap in the visible range ($E_g$~1.5 eV)[46,47] exhibits an extraordinary photoresponse to UV-visible light, making it highly promising for synaptic applications, such as a retina-like neuromorphic visual synaptic system. We investigated light-dose-dependent plasticity, encompassing variations in illumination time, optical pulse width, pulse number, pulse delay time, pulse frequency, light wavelength, and optical power. In our device measurements, we employed a 405 nm light source to stimulate the optimized device and performed all synaptic measurements, as this UV wavelength showed the best responsivity and synaptic features. As shown in Figure 3e, the EPSC (which we refer to as conductance) of the $MoWS_2$ device slowly increases from 43.8 nA to 78.4 nA ($\Delta_{EPSC}$ ~26 nA) with increasing single optical exposure period, and gradually decays thereafter once the light stimulus is removed. The decay time of the EPSC curve provides crucial information about the device's optical memory retention period, as well as quantitative information on the lifetime of the photogenerated charge carriers, which is extracted by fitting the experimental data to a double-exponential decay function. [48]

$$I(t) = I_0 + exp^{\left(\frac{-t}{\tau_1}\right)} + exp^{\left(\frac{-t}{\tau_2}\right)} \quad ........[3]$$

By fitting the long pulse potentiation data shown in Figure 3f for 80 s of light illumination using a double exponential decay function, we estimated the fast and slow decay relaxation time constants. The slow relaxation time constant $\tau_1$ (352.1 s) is found to be nearly one order of magnitude higher than the faster decay time constant $\tau_2$ (35.6 s). The dual-decay mode of the device is associated with complex carrier dynamics at the semiconductor interface and within the lattice itself. The observed slow decay time constants may be attributed to the trapping and de-trapping of photocarriers by the abundant sulfur vacancies, which generate deep and shallow-level defect states within the bandgap, in CVD-grown $MoWS_2$[48]. On the other hand, the fast decay constant is associated with direct band-to-band transition, which is related to the direct recombination of the photo-induced electron-hole pairs. The unique band structure of $MoWS_2$, with both the shallow and deep-level defect states, leads to the effective separation of the photogenerated charge carriers, thereby prolonging the photocarrier lifetime. Hence, the optical memory retention in the alloyed device is significantly improved. The decay time constants for the $MoWS_2$ alloy device are considerably higher than those of its binary counterparts, $MoS_2$ ($\tau_1$ ~147.6 s and $\tau_2$ ~19.4 s) and $WS_2$ ($\tau_1$ ~ 148.1 s and $\tau_2$ ~ 6.83 s, for identical illumination conditions (80 s, 405 nm, power 10 mW/cm$^2$). The fitted results for EPSC decay are displayed in Figure S10a. Improved retention of the $MoWS_2$ device is attributed to the higher S-vacancy (trap centre) density in the alloy, resulting from the high-temperature growth process and the possible substitution of Mo atoms into the W lattice or *vice versa*, which disrupts the crystal symmetry, thereby introducing structural disorder. The retention time can be further improved by controlling the density of the sulfur traps, which causes shallow-level defect-induced energy states in $MoWS_2$ alloy. These observed findings reveal that controlled Mo/W alloying provides a powerful means of engineering the defect landscape, thereby bridging atomic-scale lattice disorder with macroscopic neuromorphic functionality.

In biological synapses, the fundamental mechanism governing the brain's learning and memory lies in the process of STP-to-LTP transition[43,49]. This transition can be achieved by repeated rehearsal. Such plasticity can be further strengthened or weakened in our device by varying several input stimuli, such as illumination time, pulse number, optical power, and pulse rate, thereby transforming short-term memory (STM) to long-term memory (LTM). With UV light stimulation (405 nm), our device facilitates spike-duration-dependent plasticity (SDDP), as shown in Figure 3e, a continuously increasing trend in EPSC amplitude with prolonged illumination time, followed by longer retention, emulating synaptic behaviour as it transitions from STP to LTP. The observed significant conductance change, $\Delta_{EPSC}$ = 14.14 nA, for a long spike of pulse width 20 s, compared to a short spike of pulse width only 100 ms with $\Delta_{EPSC}$ = 0.7 nA, directly indicates a stronger synaptic connection via STP-to-LTP, as well as STM-to-LTM conversion. Similarly, synaptic connections can be further reinforced by increasing the pulse number, which increases the EPSC value almost linearly. In an artificial synapse, the number of optical pulses directly regulates the synaptic weight (conductance/current) modulation. The EPSC value increases continuously with a train of short-pulsed stimulation, with each pulse having a width of 500 ms and a delay of 500 ms, as shown in Figure 3g. This phenomenon replicates the spike-number-dependent plasticity (SNDP) feature of a typical artificial synapse, which is very relevant for simulating an ANN for digit and image recognition. A higher number of pulses leads to a stronger, more persistent memory. Thus, STP-to-LTP transition and linear synaptic weight update can be achieved with a higher number of optical pulses. However, as the pulse number exceeds 250, a slight non-linearity is observed, as shown in Figure 3h (peak EPSC vs pulse number). The EPSC rate decreases as the pulse number increases, with the onset of non-linearity. The pulse number determines the number of available conductance states and, hence, the regime in which in-memory computing will occur. Thus, repeated learning and relearning strengthen synaptic connections and enhance memory retention in our ternary alloy device, resembling the human brain's learning and memory behaviour. Thus, learning efficiency increases with the number of training sessions. We also observed that $MoWS_2$ OSD exhibits higher LTM relative to $MoS_2$ and $WS_2$ OSDs for the same number (N=100) of identical pulses (Figure S10b). Moreover, other light-dose-dependent properties, such as optical power, read voltage, effect of pulse width and pulse delay (each pulse with 500 ms, delay 500 ms for 10 pulses), have been investigated to emulate the synaptic transition from STP to LTP to demonstrate spike-voltage dependent plasticity, spike-width dependent plasticity (SWDP), spike-delay dependent plasticity (SDDP). The results are given in Supplementary Figure S11 (a, b and c). So, the Supplementary Figure S12 (a, b) correspond to STM to LTM conversion through spike-power-dependent plasticity (SPDP) and spike-bias-dependent plasticity (SBDP) for a single long optical pulse (pulse width=80 s), respectively. One can model the STM-to-LTM transition using Wickelgren's equation, which describes the biological forgetting law[50] given by

$$I = \lambda(1 + \beta t)^{-\psi} \qquad \text{.........[4]}$$

Where $I$ is the transient EPSC, t is the decay time, $\lambda$ is the learning state parameter at t = 0 s (degree of learning), $\beta$ is the scale parameter, and $\psi$ is the forgetting rate. The experimental EPSC curves for three distinct pulse numbers are well modelled by the Wickelgren power-law with different forgetting rates (Figure S13a). Figure S13b illustrates the relationship between memory capacity and forgetting rate as a function of training pulse number, indicating that $\psi$ (ranging from 0.56 to 0.13) diminishes markedly with an increase in spike count, whereas $\lambda$ exhibits an entirely contradictory trend. Short-term memory (STM) at lower pulse numbers produces low learning and high forgetting states, while a stronger memory is built up with higher pulse numbers, followed by higher $\lambda$ and lower $\psi$ values.

Consolidation of memory, as achieved by the increase in optical power and the subsequent STP to LTP transition, is evident in Figure 3i, with the illumination intensity varying from 1.68 $mW/cm^2$ to 11.21 $mW/cm^2$ for a train of 10 identical optical pulses (405 nm, pulse width and pulse delay, each 500 ms), resulting in an EPSC change from 35.36 to 41.60 nA. At lower power, the EPSC shows STP behaviour, whereas at higher power, both LTP and LTM are dominant, with prolonged optically induced memory retention. The peak EPSC value versus power density plot is illustrated in Figure S9b, where EPSC increases non-linearly with increasing power $((I \propto P^{\alpha}, \alpha < 1))$. The nonlinearity of the EPSC enhancement indicates the presence of defect states in the active material, leading to the capture of photogenerated carriers.

Following assumption can be considered to understand the synaptic behaviour in our $MoWS_2$ device, which promotes the photogeneration of charge carriers. Persistent photoconductivity arises from defect-mediated carrier trapping, which suppresses electron–hole recombination. Upon laser illumination, photons with energy greater than or equal to the bandgap excite electrons from the valence band to the conduction band, thereby generating electron–hole pairs in the $MoWS_2$ film. Under the influence of an externally applied bias, the photogenerated electrons in the conduction band drift toward one electrode, while the holes in the valence band migrate in the opposite direction, resulting in the generation of photocurrent. However, owing to the presence of defect-induced localized trap states within the bandgap, a fraction of the photoexcited electrons gets captured without participating in the charge transport. These defect states act as temporary carrier trap centres, effectively suppressing the direct recombination of electrons and holes. The trapped electrons can later acquire sufficient thermal energy or additional optical energy to escape from the defect states and re-enter the conduction band, where they again contribute to electrical conduction. This delayed release process significantly prolongs the effective carrier lifetime, resulting in a slower rise in the photoconductive response under continuous illumination. After the light source is switched off, electrons in the conduction band of an ideal (defect-free) $MoWS_2$ film would rapidly recombine with holes in the valence band, resulting in a fast decay of photocurrent. In contrast, in defective $MoWS_2$ films, a portion of the electrons remains trapped in vacancy-induced defect states and requires thermal activation to be released back into the conduction band before recombination. Consequently, recombination dynamics are substantially delayed, resulting in an extended photocurrent decay time and persistent photoconductivity. Alloy engineering effectively increases the retention time due to the conversion of deep-level defect states into shallow-level and the modified optical bandgap[51,52]. The prolonged decline of EPSC following extended exposure signifies enhanced light-induced memory retention. Furthermore, synaptic plasticity is significantly dependent on the frequency of the light pulses used for stimulation, a phenomenon akin to the high-pass filters seen in biological synapses[53–55].

## c) Human perceptual learning behaviour simulation:

In addition to the synaptic properties, our OSD device exhibits learning-forgetting-relearning behaviour, a characteristic of the human perceptual learning process. The human learning process consists of three parts: learning, forgetting, and relearning, as illustrated in Figure 4a. We successfully mimicked this functionality by stimulating the device with repeated pulses (pulse width ~500 ms and a 500 ms delay), which is equivalent to the learning process, then allowing the current to decay (similar to the forgetting behaviour), and subsequently stimulating it with fewer pulses to demonstrate the relearning process. This behaviour is shown in Figure 4b. In our work, we stimulated the device with a train of 29 identical optical pulses, with a maximum EPSC of 47.56 nA at the 29th pulse. The current decreases as soon as the light pulse is removed and finally returns to the initial level after 30 s, akin to the forgetting nature of the human brain. Subsequently, the same current level is achieved with only 15 pulses, resulting in a faster relearning process on rehearsal. The second EPSC exhibits a longer time course of forgetting (45 s) than the first (30 s), mimicking improved long-term memory retention after multiple learning experiences. Forgetting behaviour in our synaptic device follows the Ebbinghaus forgetting curve, with an initial forgetting rate that slows over time, akin to human perceptual learning. Figure 4c illustrates the variation of EPSC in response to 50-second-long light pulse train exposure at 405 nm, with frequencies of 0.1 to 4 Hz (each pulse has a pulse width =200 ms, $V_{read}$=0.5 V). The maximum EPSC and the decay time (forgetting time) increase with increasing frequency, thus lower frequency (0.1 Hz) induces STM, while a higher frequency (4 Hz) promotes the LTM Thus, synaptic plasticity is found to be closely related to the frequency of stimulation. The EPSC begins to diminish following the exposure to consecutive light pulses, which may be accurately modelled using the established Kohl-Rausch stretched exponential function[56].

$$I\,(t) \;=\; I_0\, exp\left[-\left(\frac{t}{\tau}\right)^{\gamma}\right] + I_{\infty} \qquad \ldots\ldots\ldots[5]$$

where, $I_0$ is the pre-exponential function, $I_\infty$ is the final value of EPSC after a certain time $t$, $\tau$ is the retention time, and $\gamma$ is the index ranging from 0 to 1. The fitted decay curves of EPSC after the application of successive 405 nm light pulses are displayed in Figure S14, and the extracted τ values for frequencies 0.5 and 2 Hz are found to be 46.69 s and 57.60 s, respectively. The increasing value of τ with frequency effectively demonstrates the STP-to-LTP transition, characterized by a much slower decay of the EPSC at higher frequencies. Figure 4d shows the EPSC amplitude corresponding to the first optical pulse ($A_1$) and the $10^{th}$ optical pulse ($A_{10}$) at varying frequencies. The EPSC gain (specified as gain = $A_{10}/A_1$) has been estimated to assess the extent of EPSC enhancement by pulses of varying frequencies. The EPSC initially rises markedly with the optical pulse frequency before approaching saturation. The EPSC gain has been modelled using a sigmoid function, a characteristic feature of the biological synapses[57,58], as follows.

$$G_f = \frac{A_1 - A_2}{\left(1+\frac{f}{f_c}\right)^n} + A_2 \qquad \ldots\ldots\ldots[6]$$

Where $G_f$ and $n$ are the gain value and order, $f$ and $f_c$ are the frequency and cutoff frequency of the input optical pulse, $A_1$ is the initial amplitude, and $A_2$ is the final amplitude. The cut-off frequency is estimated to be 0.95 Hz from fitting the curve. The above functionality of our OSD device can be promising for image sharpening applications, where the device, as a high-pass filter, can pass frequencies above $f_c$ and block others below it. Essentially, this function utilises the Fast Fourier Transform (FFT) to convert the real image from the spatial to the frequency domain, allowing only high-frequency components beyond $f_c$ to pass through, thereby sharpening the image. Thus, our CVD-grown $MoWS_2$ alloy synaptic device enables unique filtering capability for signal processing, edge detection and image sharpening applications.

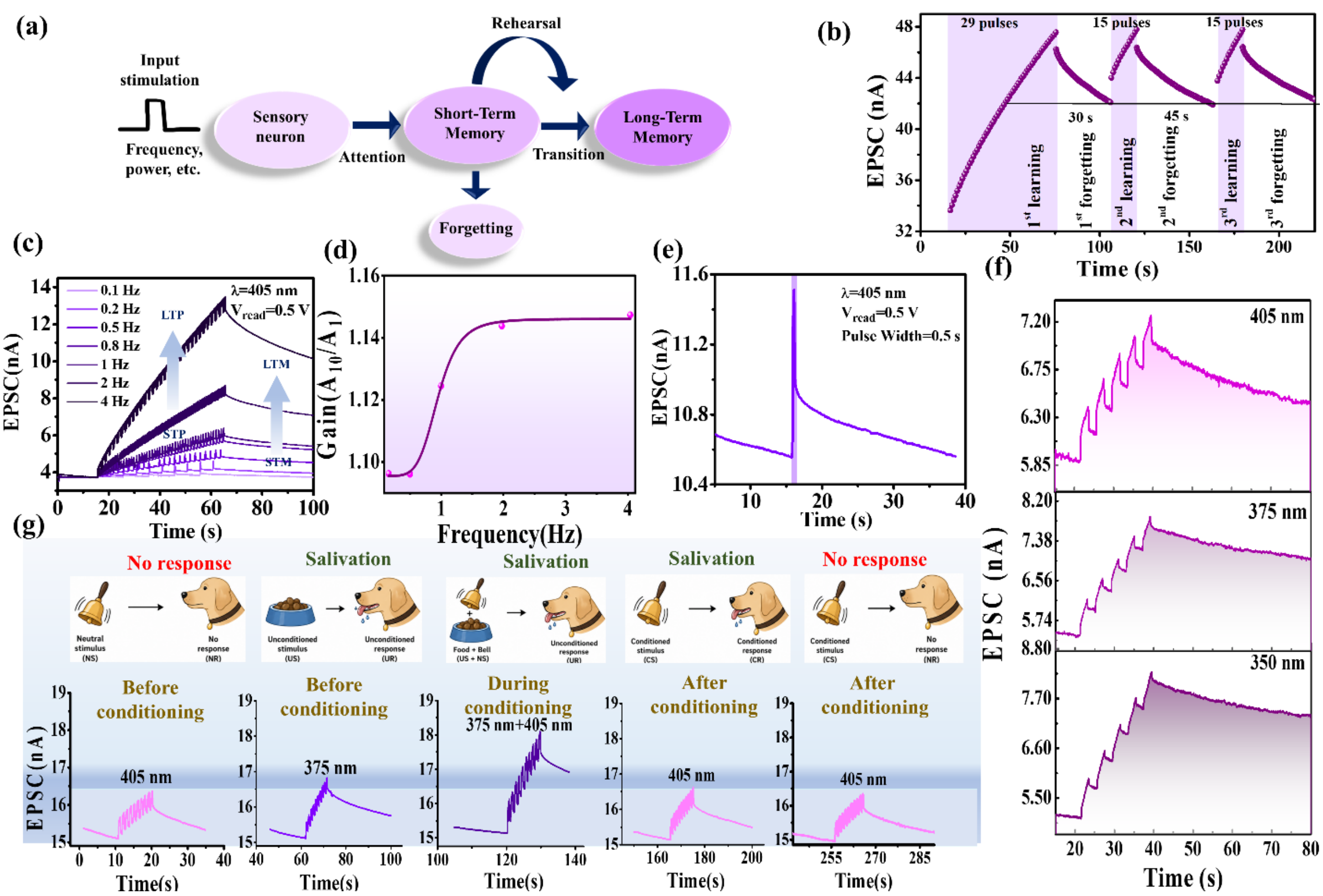

*Figure 4: (a) Schematic of human perpetual learning and forgetting process (b)learning, relearning, and forgetting behaviour realised in* $MoWS_2$ *OSD device (c) Optical spike frequency-dependent STP-to-LTP transition (d) Dependence of the EPSC gain ($A_{10}/A_1$) on the spike frequency (e) A single optical spike with PW= 500 ms, and power density= 7.64 mW/cm², operated at a bias voltage of 0.5 V (f)Spectral response of the OSD device to different UV range wavelengths.(g) Demonstration of Pavlov's dog experiment; mimicking the associative learning behaviour.*

Figure 4f demonstrates the spectral response of the synaptic device for various UV region wavelengths (350 nm, 375 nm, and 405 nm), all with the same pulse width (PW = 2s, PD= 2s, 5 pulses) and the same light intensity (~1mW), using a SQL 200 monochromator to study device response on optical stimulation of different wavelengths, synaptic potentiation effect and the conductance state modulation as evident. Our OSD exhibits enhanced synaptic properties at lower stimulation wavelengths, such as 350 nm, 375 nm, and 405 nm. The higher EPSC can be attributed to the higher absorption and the highest external quantum efficiency of $MoWS_2$ in the lower UV wavelength region, whereas at higher wavelengths, the EPSC current decreases, accompanied by a decrease in the optical memory. The wavelength-dependent spectral responses for a single short pulse (pulse width=500 ms) and a train of 100 pulses to check their spectral responsivity and potentiation phenomena for UV-green-red regions are given in Supplementary Figure S16a and b, respectively. This observation is significant because the same device can be operated in dual modes—strong synaptic and weak synaptic modes —simply by changing the input stimulation wavelength. Our retina-inspired synaptic device not only responds to light intensity and wavelength but also filters background noise, thereby enhancing contrast, improving processing efficiency, and increasing recognition accuracy in the visual cortex. Therefore, the STP-to-LTP transition with longer light-induced memory retention occurs at a lower wavelength, whereas the weak STP effect is only prominent at wavelengths beyond 405 nm. The wavelength-induced STP effect can be utilised for several purposes, including associative learning (e.g., the Pavlov dog experiment), logic circuit demonstrations, and photon-energy-sensitive pain nociceptors, etc. Consequently, a straightforward, transfer-free CVD-grown $MoWS_2$ alloy film provides a potential foundation for advancing optically stimulated synaptic devices, offering exceptional adaptability to diverse stimuli and facilitating understanding of the dynamic responses of artificial synapses to various factors. Thus, this device will serve as a channel between neuromorphic computing components and actual sensory inputs, which may be a crucial factor in advancing photonics-enabled neuromorphic networks. It is essential to minimize the energy consumption for a synaptic device to create an energy-efficient, scalable ANN. The energy consumption for optical synapse for each synaptic event can be calculated using the equation: $E = I_{EPSC} \times V_{read} \times \Delta t$, where $E$ denotes spike energy, i.e. the energy expenditure cost for each synaptic event, $V_{read}$ is the read voltage and $\Delta t$ is the duration of the optical spike. The single short spike for a synaptic event is shown in Figure 4e. In our case, the minimum energy consumption is found to be 239 pJ for a single optical synaptic training event of pulse width $\Delta t$=500 ms at a read voltage, $V_{read}$=0.5 V. This energy is 4 times lower than that of OSD made with standard Si technology-based CMOS neuromorphic device [59]. Further reduction of $V_{read}$ and $\Delta$t can help in lowering the energy requirement to train a synaptic event, similar to that used by the brain to process a single synaptic event (1 to 100 fJ)[60].

Pavlov's dog experiment is a classic example of a conditioning experiment, demonstrating the association between different stimuli — a fundamental feature of biological neural networks[4,61]. This associative learning has been demonstrated with our device using light stimulation at different wavelengths, exploiting wavelength-dependent synaptic plasticity, as shown in Figure 4f for 375 and 405 nm stimulation. IV characteristics of $MoWS_2$ OSD under dark, 375 nm and 405 nm are shown in Figure S8b. As illustrated in Figure 4g, 375 nm UV pulsed strong signals (~11 mW/cm$^2$, PW=0.5 s, 10 pulses) were used to simulate unconditioned stimuli (US) to feed the food, while 405 nm pulsed slightly weak signals (~11 mW/cm$^2$, PW= 0.5 s, 10 pulses) were used to simulate conditioned stimuli (CS) by ringing the bell. The threshold EPSC for the dog to undergo conditioned reflexes and release saliva was set at 16.36 nA. Experiments indicated that prior to training, 10 consecutive weak light stimuli failed to induce salivation, while 10 successive strong light stimuli successfully elicited it, in accordance with the defined EPSC threshold value ($I_{th}$). Consequently, the trained dog could salivate solely in response to bell stimulation following training that involved multiple bells and concurrent food signals (Figure 4g). Thus, the CVD-grown $MoWS_2$-based light-stimulated synaptic device can replicate Pavlov's classical associative conditioning experiment using two distinct light stimuli.

## d) Applications in MNIST hand-written digit recognition and color image filtering

Further, to evaluate the performance of our device in pattern recognition tasks, we implemented a neural network using the extracted device parameters as the first-layer inputs to classify the MNIST handwritten-digit dataset and quantified how retention affects performance over time. Figure 5(a) summarizes the implemented multilayer perceptron (feedforward artificial neural network) used for hand-written digit classification. The 28×28 grayscale input image of a digit is flattened into a 784-dimensional vector, projected to a 256-neuron hidden layer, and mapped to 10 output classes. In the code, the synaptic weights of the first matrix (784→256) are mapped to the measured device conductance states. A central step in implementing the ANN is quantizing synaptic weights and biases to the experimentally available conductance levels. Figure 5b shows the positive potentiation for 100 consecutive identical optical pulses (PW=500ms, 0.5 V, 405 nm). It is representative of the EPSC vs time curve for 100 pulses and the subsequent decay behaviour. The conductance weight update increases monotonically with pulse number, increasing from 1 to 100 (Figure 5c), illustrating a significant requirement for artificial neural network applications, such as hand-written digit recognition. The graph has been fitted with a non-linear fitting equation ($G \propto P^{\nabla}$), with $\nabla$ being the non-linearity factor, $\nabla = 0.73819$, evaluated from the fitting results. Using the device's potentiation curve, we construct a discrete set of allowed conductance values using [62]

$$G_p = G_{min} + (G_{max} - G_{min})\left(\frac{1-e^{-\frac{\nabla p}{100}}}{1-e^{-\nabla}}\right) \quad \text{.........[7]}$$

where $G_p$ is the output conductance weight update, $G_{min}$ *and* $G_{max}$ are the minimum and maximum conductance values, $\nabla$ is the non-linearity factor, and $p$ is the total number of levels (100) in our device measurements. The trained weight matrix is then projected onto this discrete set (nearest-level assignment) following the differential encoding scheme[63,64], so that every synapse corresponds to an achievable conductance state. In our case, $G_{min}$=27.9719 nS, $G_{max}$=73.0092 nS, $p$=100 and ∇=0.73819. The non-linearity factor ∇ has been derived by fitting the conductance vs pulse number curve using the equation [7]. Figure 5d reports the device accuracy with increasing epochs as the solution is iteratively refined under the positivity and conductance-range constraints, rising rapidly and saturating close to 92.43%.

To further test the reliability of our device for classification algorithms, we incorporate the device's decay characteristics into the network model. Initially, we store the conductance data following the above-described procedure and then, for studying the effect of its decay over time, we calculate the modified memory retention using the Kohl-Rausch stretched exponential at each time step and apply the forward pass with the modified values. The Kohl-Rausch exponential function is fitted to the decay characteristics of a 50-pulse data set (from Figure 3g), corresponding to the mid-potentiation condition. This state was chosen deliberately as the average case: since the 100 conductance levels used in the network are in the 0-to-100-pulse range, statistically half of them lie below the peak conductance reached at 50 pulses and half lie above it, making this the most representative single reference for decay behaviour across the full range of levels, given that retention itself depends on how potentiated the device is.

The Kohl-Rausch stretched exponential function[65] is given by the equation

$$M_t = A \exp\left[-\left(\frac{t}{\tau_r}\right)^{\alpha}\right]..................[8]$$

where A is a constant, $\tau_r$ is the characteristic retention time, and α is the stretch index, ranging between 0 and 1, and $M_t$ is the memory retention percentage defined as:

$$M_t = \frac{I_t - I_{dark}}{I_{max} - I_{dark}} \times 100\%.................[9]$$

Where $I_t$ is the current changing with time, $I_{dark}$ is the dark current without any light, and $I_{max}$ is the maximum current before the termination of illumination.

The immediate outcome is that the device-mapped network reaches high classification performance initially. However, the reliability picture changes when the same conductance-mapped network is evaluated at increasing times after programming. Figure 5e plots accuracy versus time. The accuracy remains high initially but then exhibits a pronounced decline beyond the characteristic timescale of conductance relaxation, ultimately approaching near-chance performance (~10%, precisely 8.92% in our case) at longer times.

Fitting the decay characteristics from the 50-pulse (average-case) data using Equation [8], we find the memory retained in the device to decrease relatively faster with time, dropping to near-initial values (Figure S17a). Despite this, the trained neural network's accuracy (Figure 5e) remains comparatively stable for a much longer time, staying close to its initial high value before showing a markedly delayed degradation compared to the memory retention of our device, indicating that the classification performance of the device-mapped network considerably outlasts the retention of the raw conductance (or memory) values themselves.

The confusion matrices provide a diagnostic view of the failure mode. Figure 5f shows the normalized confusion matrix at the initial evaluation time: the strong diagonal values confirm robust class separability with only minor misclassifications. In contrast, at the final time point (Figure S17b, after 10,000 s), the predictions collapse predominantly into a single output class (a near-unit column), indicating a catastrophic failure rather than a gradual, uniformly distributed misclassification. This behaviour is consistent with drift-induced suppression and homogenization of the first-layer conductance states, after which the network’s downstream decision becomes dominated by fixed non-memristive offsets and biases, yielding a trivial classifier. Taken together, Figures 5 (a-f) support two conclusions. First, the hardware mapping confirms that the fabricated device can implement quantized, non-negative synaptic weights for a practical inference task, reaching high accuracy shortly after programming (Figure 5d and f). Second, despite the rapid decay of memory (or conductance) predicted by the adopted Kohlrausch drift model, the neural network retains high inference accuracy for significantly longer periods. This striking disparity between device-level degradation and algorithm-level performance reveals unexpected robustness in the system, demonstrating that useful computational functionality can persist long after substantial conductance relaxation.

A key feature of the human retina is its extraordinary ability to discriminate colors. To emulate the retina's color discrimination capability in our device, red (630 nm), green (525 nm), and blue (405 nm) light stimuli were applied to the $MoWS_2$ OSD. The wavelength dependence of the synaptic response ($\Delta_{EPSC}$) was subsequently investigated, as shown in Figure S16. The distinct $\Delta_{EPSC}$ values observed across different light wavelengths arise from the wavelength-dependent optical absorption properties of $MoWS_2$. Consequently, the device exhibits wavelength-sensitive synaptic plasticity, including STP, LTP, and the transition from STP to LTP. Such spectrally selective synaptic behaviour endows the device with the capability to differentiate colors, making it promising for artificial vision applications. The wavelength-selective photoresponse of the synaptic device also enables a qualitative demonstration of spectral image filtering. Since the device exhibits a dominant conductance modulation under blue illumination, each pixel of an input color image can be mapped to a normalized conductance value using the experimentally calibrated response functions[2] for the red, green, and blue channels. Figure 5h shows the spectral response of the device for 405 nm, 525 nm, and 630 nm illumination for 10, 50, and 100 pulses. Power-dependent enhancements of EPSC with different pulse number for these wavelengths are shown in Figure S15 (a, b and c for 405 nm, 525 nm, and 630 nm). Figure 5i shows the original image containing blue flowers against a warm-toned background.

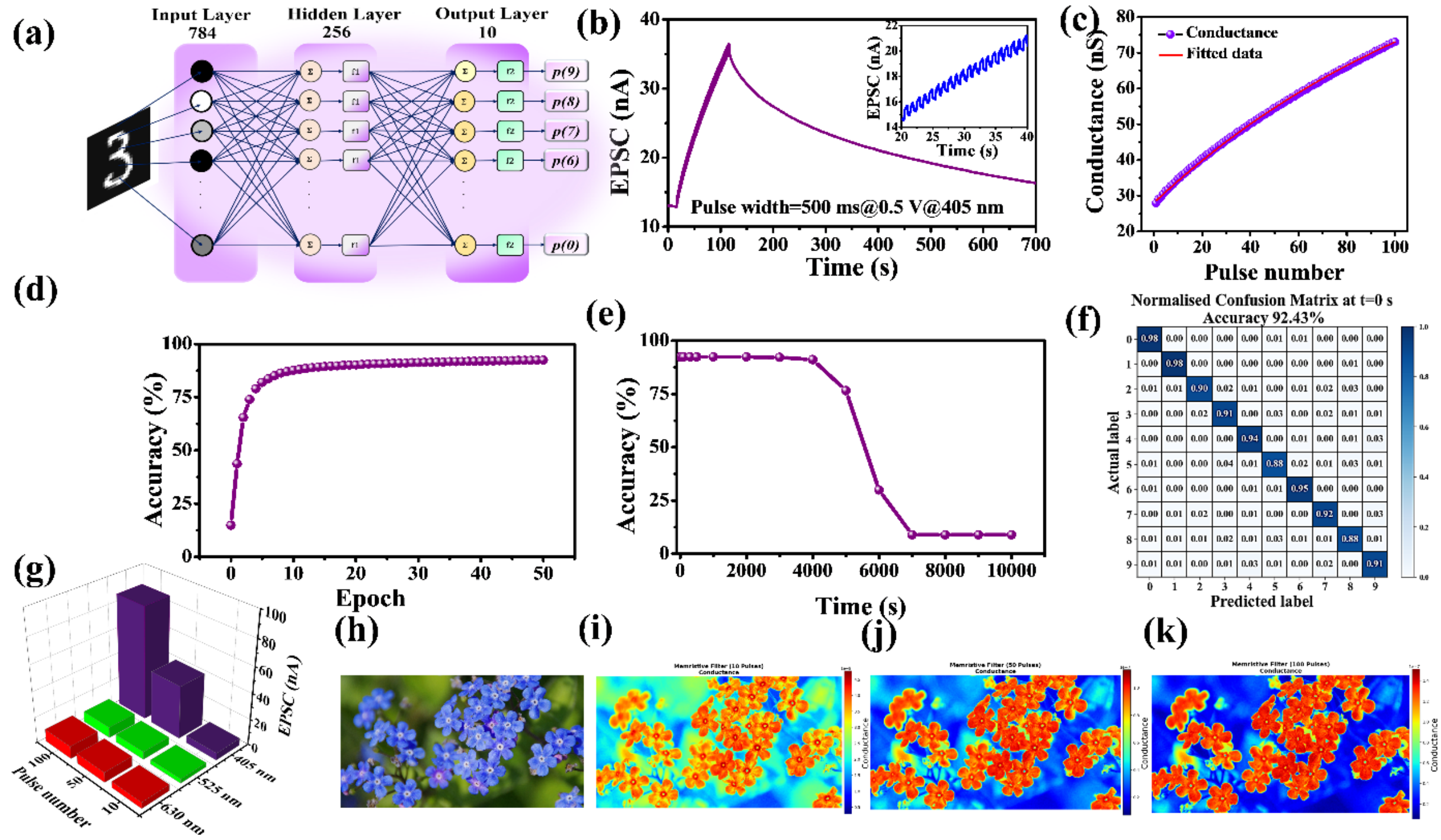

*Figure 5: (a) Schematic of artificial neural network simulation for digit recognition task (b) EPSC vs time for 100 pulses (PD & PW=500 ms, with inset showing the zoom-in image of the consecutive pulses.(c) Conductance weight update with increasing pulse number (d). Simulated device accuracy with epoch (e) accuracy drop over time, once 100 identical pulses are applied to train the model (f) Normalized confusion matrix at t=0 s with 100 pulses applied for training the device. (g) Spectral response of the OSD device at 405nm, 525 nm and 630 nm stimulation for 10, 50 and 100 pulses (h-k) Color image filtering applications using spectral dependence data, where (h) shows the original flower image and (i,k and k )stand for normalized conductance map of the flower image after 10, 50 and 100 pulses of red, blue and green illumination, respectively.*

When processed through the device response model after 10 and 50 optical pulses (Figures 5i and j), the blue-rich flower regions begin to stand out with higher normalized conductance (blue in the color map) relative to the greenish background. After 100 pulses (Figure 5k), the image contrast sharpens further as the blue-channel conductance grows more rapidly than the other channels, enabling clearer segmentation of the flowers from the background. This illustrates how the device's intrinsic spectral selectivity can serve as a hardware-level color filter, preferentially amplifying blue content without explicit digital post-processing.

From the results discussed earlier, the 2D $MoWS_2$ alloy with a high density of sulfur vacancies can contribute to synaptic functionality through defect-assisted trapping & de-trapping process of photogenerated charge carriers. Although vacancy-induced intrinsic defects can reduce carrier transport and hinder high-speed transistor performance, defect-mediated charge trapping and release characteristics are highly beneficial for analog neuromorphic memory and synaptic device applications. The higher the sulfur vacancies, the stronger the n-type doping effect, with the trap states acting as donor-like states [65]. The presence of defect states has been probed by KPFM, as shown in the surface potential mapping images (Figure 2g and Supplementary Figure S7), which reveal a highly non-uniform potential distribution, attributed to different electron- and hole trapping sites in the sample. The role of these traps and KPFM analysis has been comprehensively explained in the Supplementary Note 3. The thermally stable $Mo_xW_{1-x}S_2$ alloy exhibits a composition-dependent bandgap, distinct from those of its binary components and possesses shallow defect states [52] within the bandgap of $MoWS_2$, caused by S vacancies. The persistent photoconductivity (PPC) observed in $MoWS_2$ originates from defect- and interface-induced trap states within the bandgap. Upon illumination, electrons are excited from the valence band ($E_V$) to the conduction band ($E_C$), generating electron–hole pairs. While some carriers recombine directly, others are trapped in shallow and deep defect states. These trapped electrons and holes require thermal activation to return to the conduction or valence band, respectively, before recombination can occur. The delayed trapping–detrapping process suppresses carrier recombination, thus prolonging carrier lifetime, and gives rise to PPC[44,45]. The decay dynamics of the photoconductance can be described by the relation $t_{\text{decay}} = t_r + t_t(1 + r)$, where $t_r$corresponds to the direct carrier recombination time associated with the fast relaxation process, $t_t$ represents the carrier de-trapping time governed by the thermally activated release of trapped carriers from defect states, and $r$ denotes the probability of carriers being re-trapped prior to recombination. The de-trapping process is strongly influenced by the energy barrier ($\Delta E$) between the localized trap states and the conduction or valence band energies, following the thermal activation expression $t_t = t_0 \exp(E/kT)$[66], where $t_0$ is the characteristic time constant, $k$ is the Boltzmann constant, and $T$ is the absolute temperature. As a result, the existence of S-vacancy-induced localized trap states in $MoWS_2$ substantially prolongs the carrier relaxation dynamics by delaying recombination through repeated trapping and de-trapping processes, thereby giving rise to pronounced persistent photoconductivity behaviour. Thus, the intermediate composition (*x=0.5*) offers robust neuroplasticity and exhibits higher trap density, better memory retention, STP and LTP, effective human learning-forgetting-relearning, associative learning, and wideband spectral-responsive visual processing power.

## Conclusions:

In summary, we have successfully developed a two-terminal optical synaptic device utilizing ternary $MoWS_2$ TMD alloy layer grown via microcavity-assisted CVD reactor. Driven by growth-induced intrinsic sulfur vacancies, as verified through XPS, KPFM and STEM studies, the device successfully emulates comprehensive biological brain functionalities. The device demonstrates essential synaptic features, including excitatory postsynaptic current (EPSC), paired-pulse facilitation (PPF), and the transition from short-term plasticity (STP) to long-term plasticity (LTP). Furthermore, the OSD exhibits advanced, stimulus-dependent learning dynamics governed by spike-rate (SRDP), spike-frequency (SFDP), spike-number (SNDP), and spike-duration (SDDP) plasticities, alongside characteristic biological learning and forgetting profiles. Operating with an ultra-low energy consumption of ~239 pJ per training event, the device offers a broad spectral response with wavelength-selective operation, enabling advanced cognitive capabilities such as associative learning. At the system level, the measured synaptic conductance updates are integrated into an ANN simulation, yielding a high recognition accuracy of 92.43% on the standard MNIST handwritten digit database. Despite time-dependent decay of all synaptic conductances, the trained neural network retains inference accuracy over an extended period, indicating system-level robustness against device retention loss. Furthermore, the device’s wavelength-dependent photoresponse is successfully exploited to demonstrate hardware-based color image filtering. These findings underscore the vast potential of defect-engineered 2D ternary alloys for energy-efficient, multi-functional neuromorphic computing and visual pre-processing systems.

# Experimental Methods

## Microcavity-assisted space-confined growth of $Mo_xW_{1-x}S_2$ alloy

$MoWS_2$ alloy was synthesized using $MoO_3$, $WO_3$, and S powders as precursors for Mo, W, and S, respectively, with NaCl (all procured from Sigma-Aldrich, 99.97%) as a growth promoter. $MoWS_2$ growth was carried out in a low-pressure, two-zone CVD system (Ants Innovation) employing a space-confined microcavity reactor (Figure 1a). The microreactor is formed by stacking two cleaned $SiO_2$ substrates (1 × 1 cm², 300 nm oxide), polished surfaces face-to-face with micrometre-order air gap and placing them downstream above a quartz boat containing the precursors. $MoO_3$ (upstream) and a $WO_3$ + NaCl mixture (downstream) were loaded in a single quartz boat with a separation of ~10 cm, while sulfur was placed in a separate low-temperature zone ~30 cm away upstream. Prior to growth, the chamber was purged with Ar (500 SCCM, 15 min). During deposition, Ar carrier gas (50 SCCM) was used to transport sulfur vapor into the confined growth region, where it reacted with vapors of Mo- and W-containing species. The sulfur and metal precursor zones were heated to 300°C (ramp rate 12º C/min) and 950°C (ramp rate 10º C/min), respectively, and maintained at these target temperatures for 15 minutes to facilitate nucleation and alloy growth, leading to lateral expansion and the formation of a uniform large-area thin film under optimized conditions (Furnace temperature ramp and hold time profile is shown in Figure S1). The sulfur precursor zone heating was started when the Mo/W precursor zone reached 700ºC, well before its target temperature. Growth temperature and carrier gas (Ar) flow rate were systematically optimized to achieve uniform $MoWS_2$ films, evolved from isolated triangular flakes to continuous coverage on $SiO_2$/Si substrates (Figure S2).

## Characterization of the as-grown $MoWS_2$ alloy flake and film

CVD-grown individual flakes and films were primarily characterized immediately after growth using an optical microscope with a 100x objective lens (Zeiss). Micro-Raman and photoluminescence measurements were conducted for the characterisation of vibrational and optical properties using a confocal Raman-PL setup (WiTech Alpha 300R) equipped with a CCD detector, using a 532 nm solid-state laser. Flake and film thickness were measured using an atomic force microscope (Agilent), while high-resolution microscopic images of the film were taken in a field-emission scanning electron microscope (Zeiss VP 300). X-ray diffraction spectrum was acquired using a Beamline Synchrotron X-ray diffraction system, BL 18B, Photon Factory, KEK, Japan. The stoichiometry of the elemental composition of the $MoWS_2$ film was estimated from the X-ray photoelectron spectroscopy (XPS) system (Versa Probe II). High-resolution TEM imaging was performed with a JEOL microscope, while topographic height profiles and surface potential maps were obtained with an advanced AFM system from Asylum Research (Oxford Instruments) and a Bruker Multi-Mode AFM system, respectively.

## Optical synaptic device fabrication and characterization

We fabricated two-terminal synaptic devices using CVD-grown $MoWS_2$ films via shadow-masking with Cr/Au (5 nm/50 nm) electrodes, deposited by thermal and e-beam evaporation, respectively, with a channel spacing of 100 microns. A Keithley 4200A SCS semiconductor parameter analyzer was used for conductivity measurements under dark and illumination conditions, with different LEDs (Thorlabs) as the optical sources.

## Neural Network Simulations

The 784-256-10 ReLU network was trained on MNIST with weights encoded as differential conductance pairs, $W = G^+ - G^-$, to reflect the positive-only nature of synaptic devices. The effective weight W was optimized directly under a hard clamp, then reconstructed into a balanced conductance pair and quantized to discrete device levels. ReLU was chosen over sigmoid for its homogeneity, which makes classification robust to uniform conductance attenuation. A subsequent drift study was done to simulate the decaying nature of the conductances using the Kohl-Rausch exponential to both conductance rails and tracked test accuracy and confusion matrices across increasing drift times to characterize retention.

## Supplementary Information

Supplementary Information file is attached herewith.

## Acknowledgements

D.K.S. acknowledges the Prime Ministers' Research Fellowship (PMRF) Fellowship program, Ministry of Education, Govt. of India, for providing him with financial assistance to carry out the research work. S.K.R. duly acknowledges the financial support from DST Nanomission (DST/NM/TUE/QM8 2019G 5), and DST/QTC/NQM/QMD/2024/4(G). The authors also acknowledge Dr. Shaona Bose for her help in synchrotron XRD characterization of the alloy and binary film. The financial support by the Department of Science and Technology, India, for the experiments at the Indian Beamline, PF, KEK, Japan, is gratefully acknowledged.

## Author Contributions

D.K.S. conceived the study and designed the experiments. D.K.S. optimized the CVD growth parameters for the growth of MoWS2 alloy, performed materials characterization and analyzed the data. D.K.S. and S.D. fabricated the device by thermal and e-beam evaporation with shadow masking. D.K.S. and S.K.G. performed the optical synaptic device measurement and discussed the results. S.G. performed ANN simulations for MNIST-based digit recognition and a color image filtering application using spectral response data and discussed the simulation results. D.K.S. wrote the manuscript with input from all authors. S.K.R. formally investigated, reviewed the manuscript, validated the data, provided resources and supervised the project

## Conflict of Interest

The authors declare no conflict of interest.

## Data Availability Statement

The data that support the findings of this study are available from the corresponding author upon reasonable request.

## Keywords

## Table of Contents (ToC) image

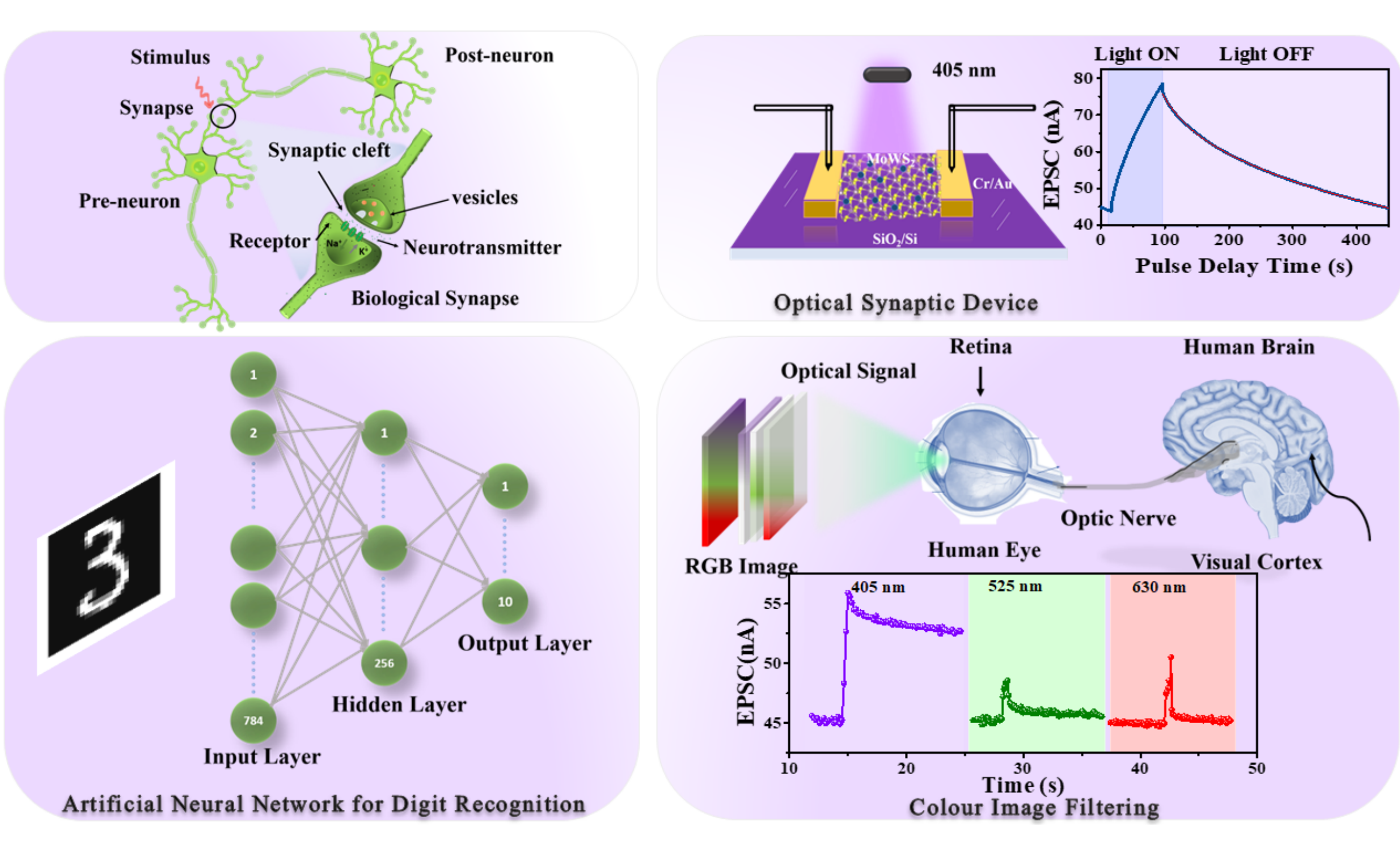

# Supplementary Information

## Energy-efficient, reconfigurable optoelectronic artificial synapses based on $MoWS_2$ alloy for pattern recognition and color image filtering applications

Deepak Kumar Sahu[1], Santu Kumar Ghosh[1], Sagarneel Ghoshal[1], , Saranya Das[1], Samit K. Ray[1,*]

[1]*Department of Physics, Indian Institute of Technology Kharagpur, Kharagpur, India, 721302*

[*]*Corresponding author email ID: physkr@phy.iitkgp.ac.in*

## Note 1: Growth mechanism of space-confined microcavity reaction in CVD:

In the case of space-confined growth, all contributing precursor vapours have sufficient time inside the air-gap-sandwiched, face-to-face-oriented $SiO_2$ substrate pairs, altering the growth kinetics from surface-reaction (edge-attachment) to mass-transport (diffusion)-limited growth, enabled by reduced gas flow velocity in the confined space regime, which resulted in large-area growth of $MoWS_2$ alloy[1]. Post-growth $MoWS_2$ grains/films were primarily identified by rendering color contrast in the optical micrographs of the grown flakes relative to the substrate, as shown in Figure 1d. The higher growth rate in the space-confined regime is attributed to differences in mass-transport rates of precursor species across the boundary layer within the quartz tube and in the vicinity of the substrate surface.

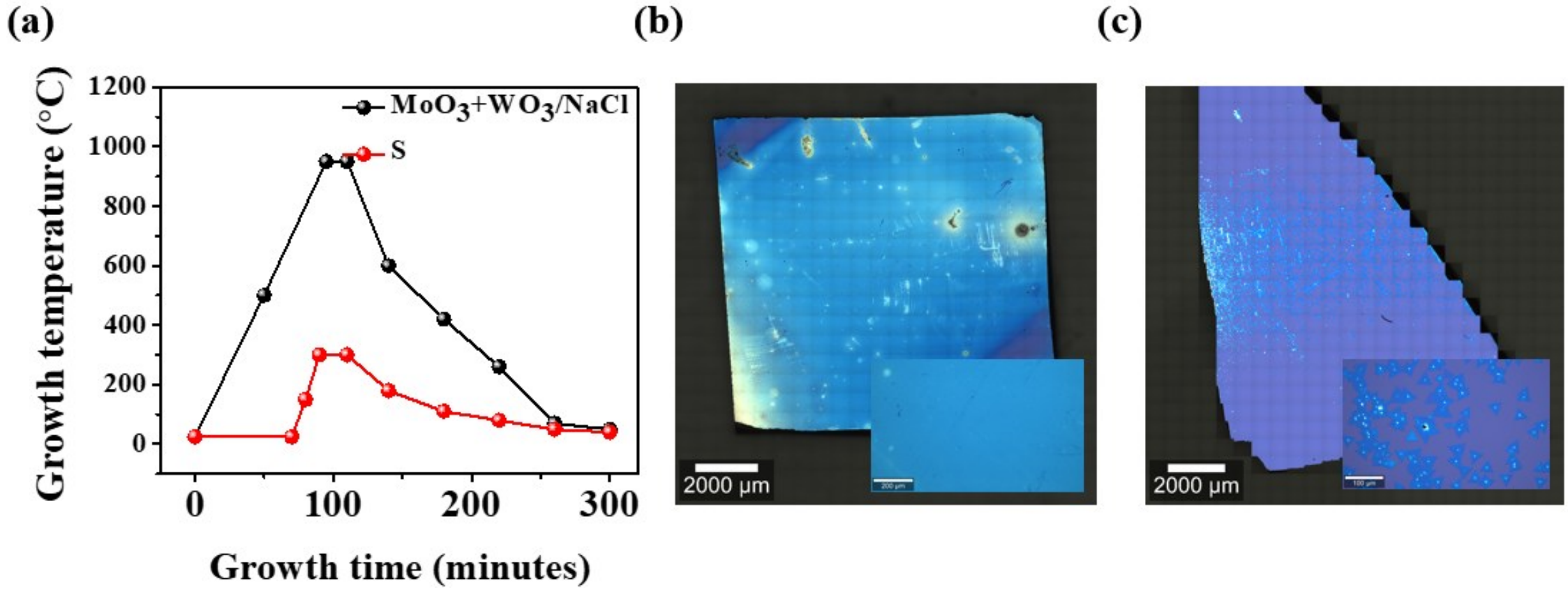

*Figure S1: Growth temperature profile of the microcavity CVD growth of $MoWS_2$. Stiched optical microscopy image of (b) thin film grown in microcavity reactor chamber; inset shows zoom-in picture of the thin film and (c) conventional substrate upside down direct growth with formation of individual flakes with the same growth conditions, as shown in the inset.*

The striking result can be explained by a typical kinetic model for thin-film deposition in a CVD reaction, where growth kinetics are determined by the competition between mass transport (diffusion) and surface reaction (edge attachment). Assuming steady-state gas flow inside the growth chamber and a stationary boundary layer with a thickness $\delta$, the following processes occur sequentially within the growth chamber: precursor diffusion into the boundary layer, followed by adsorption, and then chemical reaction, which forms the material on the substrate surface via diffusion. Depending on the gas flow near the boundary layer or the substrate surface, the two gas flow regions can be classified as a mass-transport region and an exponentially substrate-temperature-dependent surface-reaction region. Hence, there are two fluxes of the active species near these regimes, given by the following reactions.

$$F_{mass-transport} = h_g(C_g = C_s) \text{.........[S1]}$$

$$F_{surface\ raction} = K_s C_s \text{ .........[S2]}$$

here $F_{mass-transport}$ is the flux of active species through the boundary layer, $F_{surface\ raction}$ is the flux of consumed active species at the surface (assuming first-order kinetics), $h_g$ is the mass transport coefficient, $K_s$ is the surface reaction constant (assuming first-order kinetics), $C_g$ is the concentration of gas in the bulk, and $C_s$ is the concentration of the active species at the surface. These fluxes are in series, and the slower of the two processes is the rate-limiting step during TMD synthesis. At the equilibrium state, $F_{mass\text{-}transport} = F_{surface\text{-}reaction} = F_{total\text{-}flux}$, and the total flux ($F_{total\text{-}flux}$) after eliminating $C_s$ can be rewritten as $[K_s h_g/(K_s + h_g)]C_g$. Mathematically, three regimes arise: $h_g \gg K_s$ (surface reaction-controlled region), $h_g \sim K_s$ (mixed region), and $h_g \ll K_s$ (mass transport limited region). At high temperatures, mass transport through the boundary layer is rate-limiting ($K_s \gg h_g$). [2] Additionally, $h_g$ is correlated to the boundary layer thickness by the relation

$$h_g = \frac{D_g}{\delta} \text{.........[S3]}$$

The smaller the $h_g$ value, the more stable the growth environment is. Additionally, the boundary layer thickness δ is inversely proportional to the square root of the airflow velocity, as predicted by the Blasius model[31].

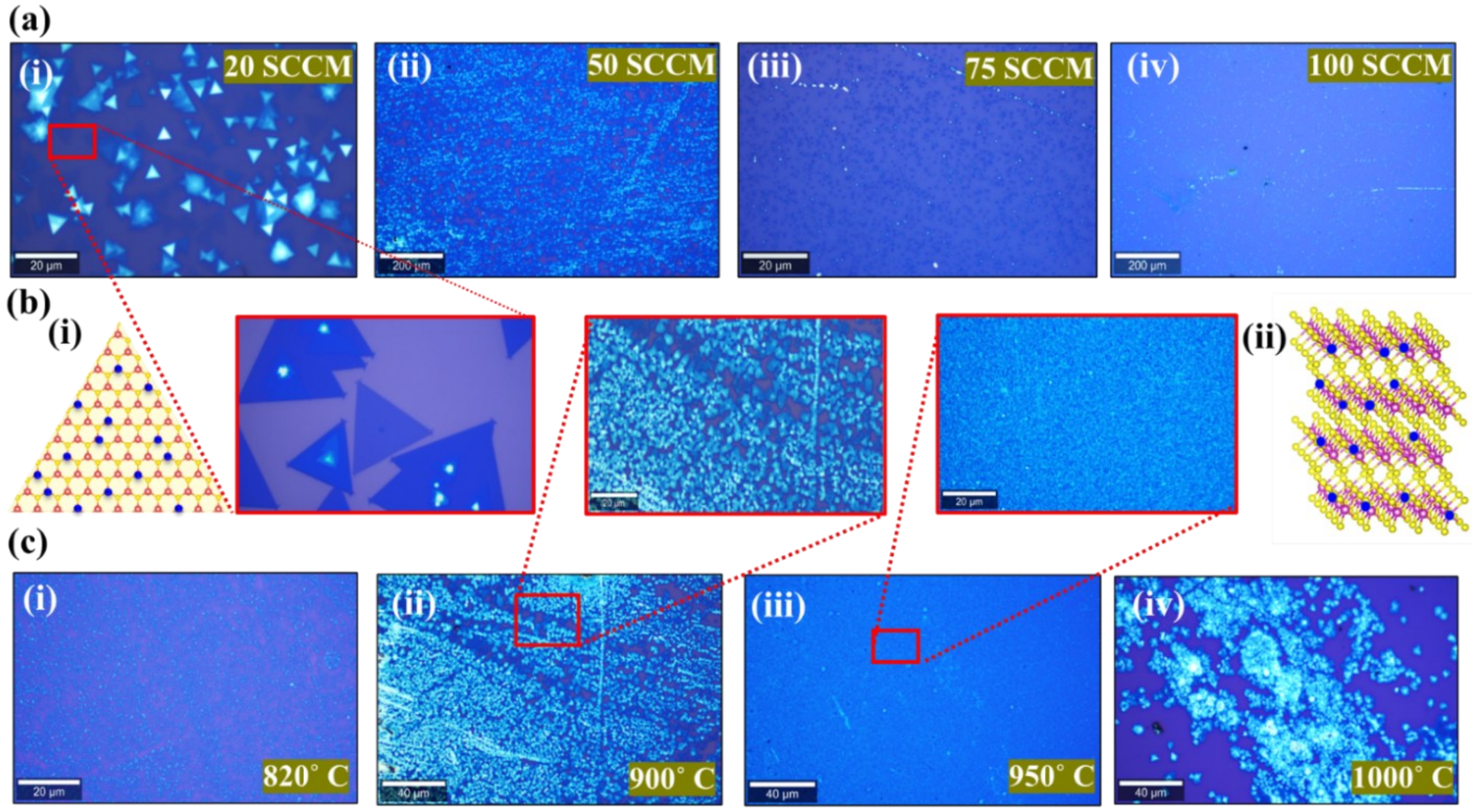

*Figure S2: Optical microscopy images of the as-grown CVD alloy $MoWS_2$ film at different carrier gas flow rates(a) (i) 25 SCCM, (ii) 50 SCCM, (iii) 75 SCCM, (iv) 100 SCCM, (b) (i) schematic of molecular structure of monolayer and (ii) few-layer $MoWS_2$ random alloy; growth pattern of the as-grown film of $MoWS_2$ on the $SiO_2$/Si substrate at different growth temperatures(c) (i) 900˚C, (ii) 925˚C, (iii) 950˚C and (iv) 1000˚C. The zoom-in pictures of the optical images shown in Figures (a, i), (c, i and ii) are shown in the middle of the Figure, marked with red blocks.*

We studied the influence of 4 different flow rates (20, 50, 75, and 100 SCCM) and observed that at a slower flow rate (20 SCCM), individual flakes were formed, whereas at a higher flow rate (more than 75 SCCM), small individual nucleation centres and average-sized flakes were visible again. An intermediate flow rate (50 SCCM) produced an optimized thin film, as shown in Figure S2a, (ii), achieved through the coalescence of several individual grains, rather than several small grains, governed by equilibrium growth kinetics. We also investigated the effect of growth temperature on sample quality by varying the temperature from 900 °C to 1000 °C in 25 °C increments. We observed better film quality and coverage at an intermediate temperature of 950 °C than at the two extreme temperatures (900 °C and 1000 °C) at a pressure of 0.5 mbar, as shown in Figure S 2c (iii).

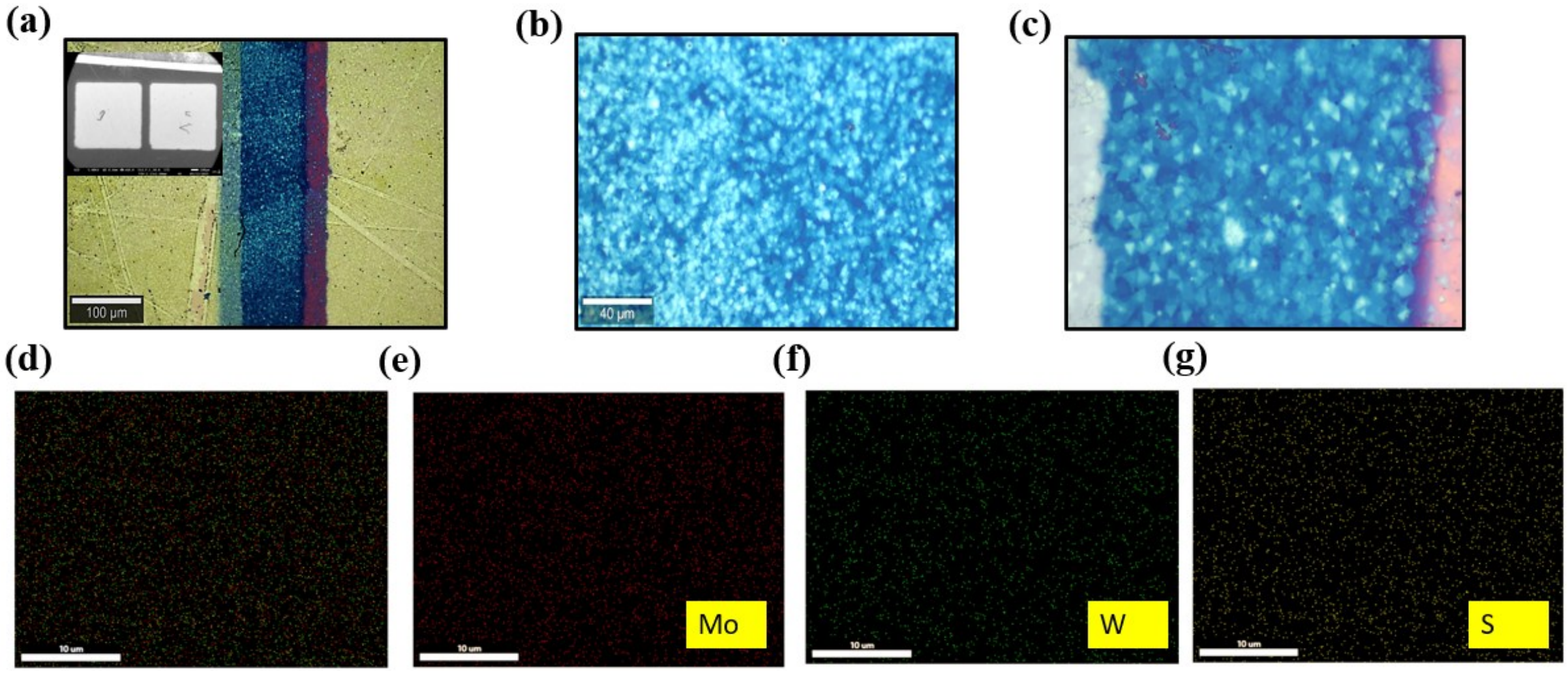

*Figure S3: (a) Optical image of the two-terminal MSM type device with a bluish $MoWS_2$ channel of 100 μm channel length; the inset shows FESEM image of the as-fabricated device (b) Optical image of the alloy film. (c) Close view of the film with distinctly visible small $MoWS_2$ flakes overlapping with each other. (d, e, f and g) EDX color mapping of the as-grown $MoWS_2$; all the elements are evenly distributed.*

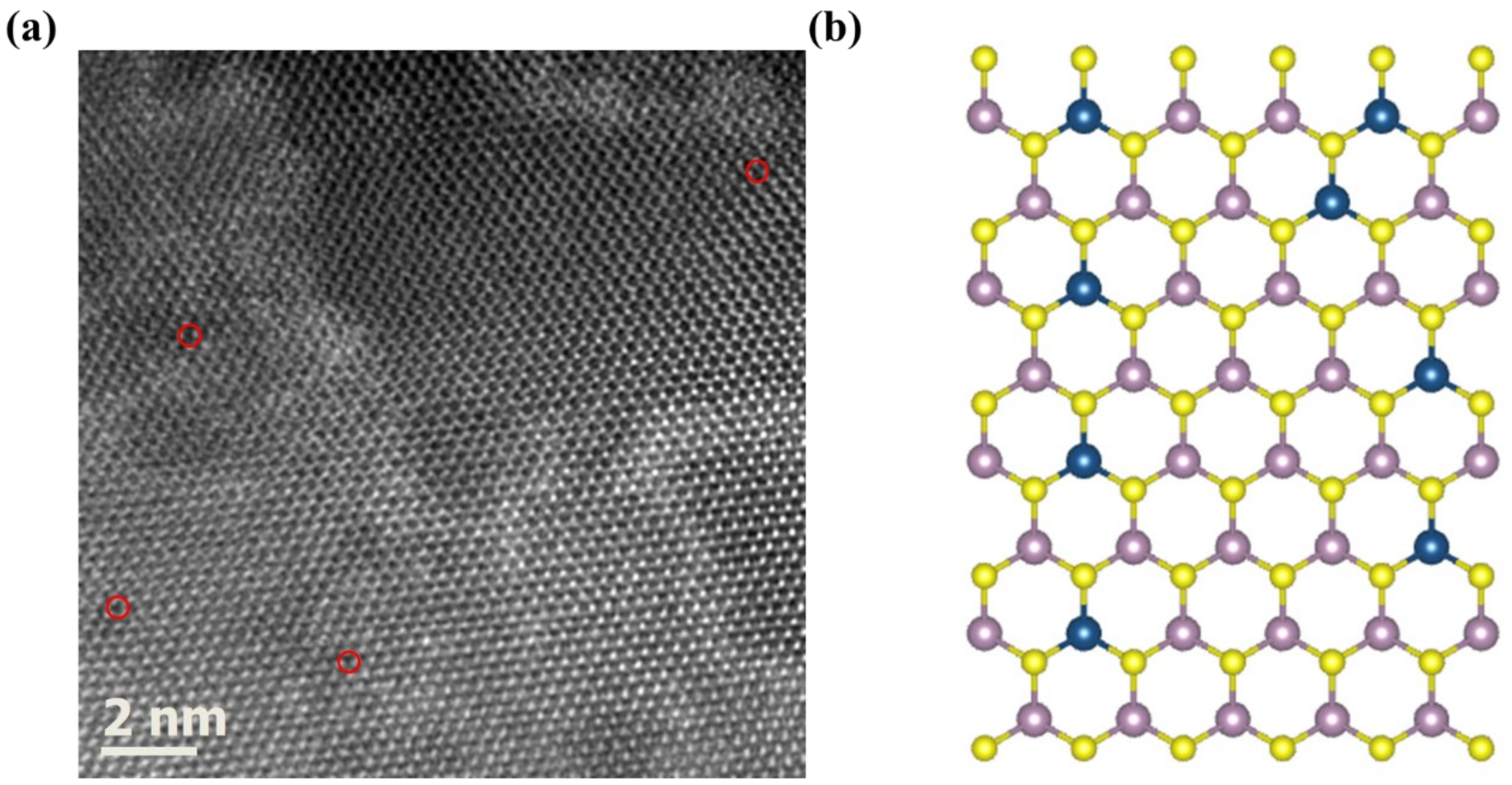

*Figure S4: (a) STEM image of $MoWS_2$ alloy indicating chalcogen vacancies labelled as red circles, the bright spots indicate the transition metal elements (more bright-W, intermediate bright-Mo), while the less-bright spots indicate the chalcogen elements. (b) Schematic model of the hexagonal pattern of the monolayer $MoWS_2$; the yellow circles indicate the sulfur atoms, purple circles for the Mo atom and the blue circle for the W atom*

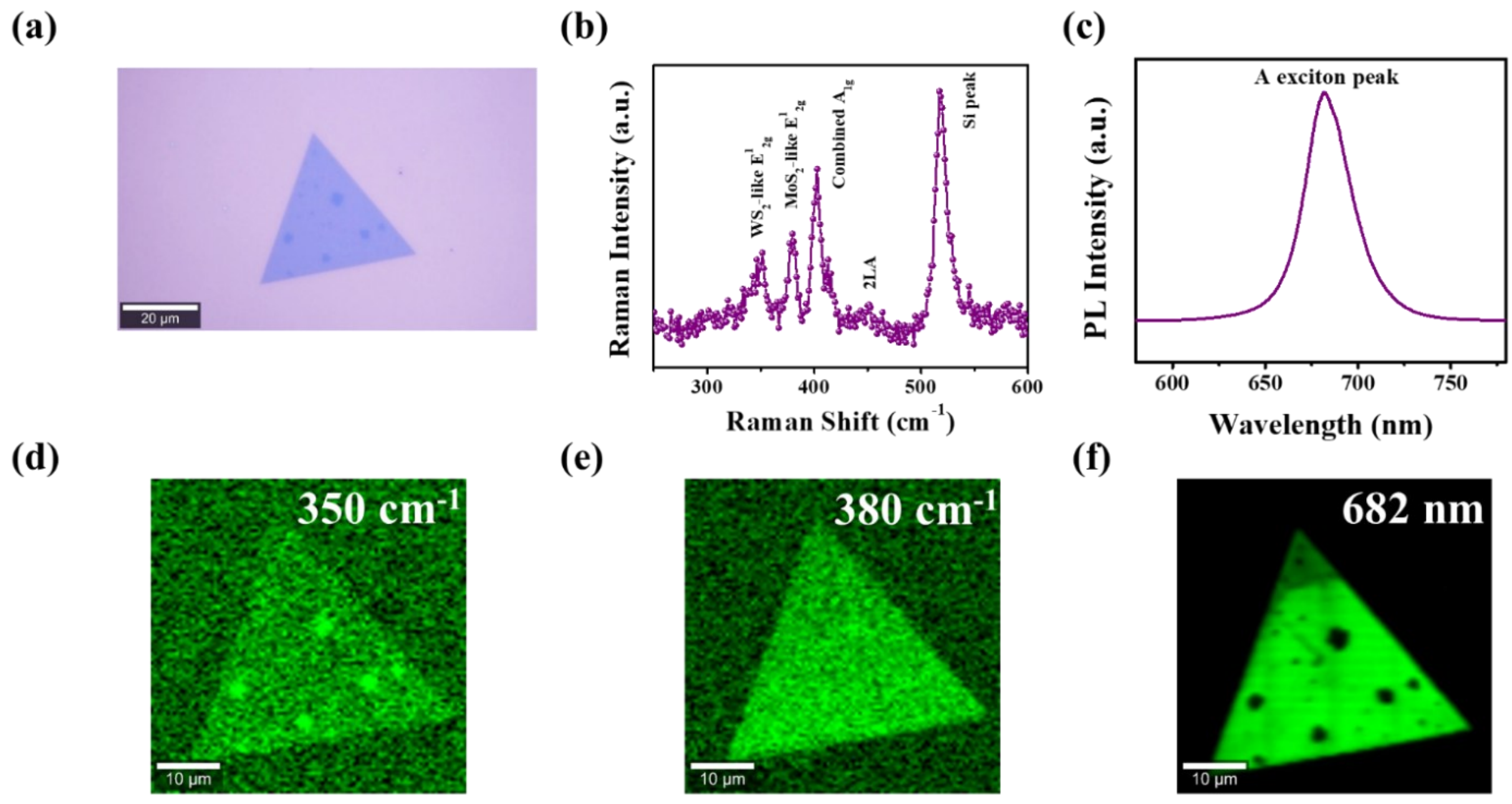

*Figure S5: (a) Optical image of a single triangular monolayer $MoWS_2$ flake. (b) Raman spectrum of $MoWS_2$ showing three characteristic peaks ($WS_2$-like $E^1_{2g}$, $MoS_2$-like $E_{12g}$ and combined $A_{1g}$ vibrational modes(c)PL spectrum of monolayer $MoWS_2$ with the prominent A excitonic at ~682 nm. (d)Raman intensity mapping of $WS_2$-like $E^1_{2g}$ Raman mode (e) Raman intensity mapping of $MoS_2$-like $E_{12g}$ (f) PL mapping of monolayer flake*

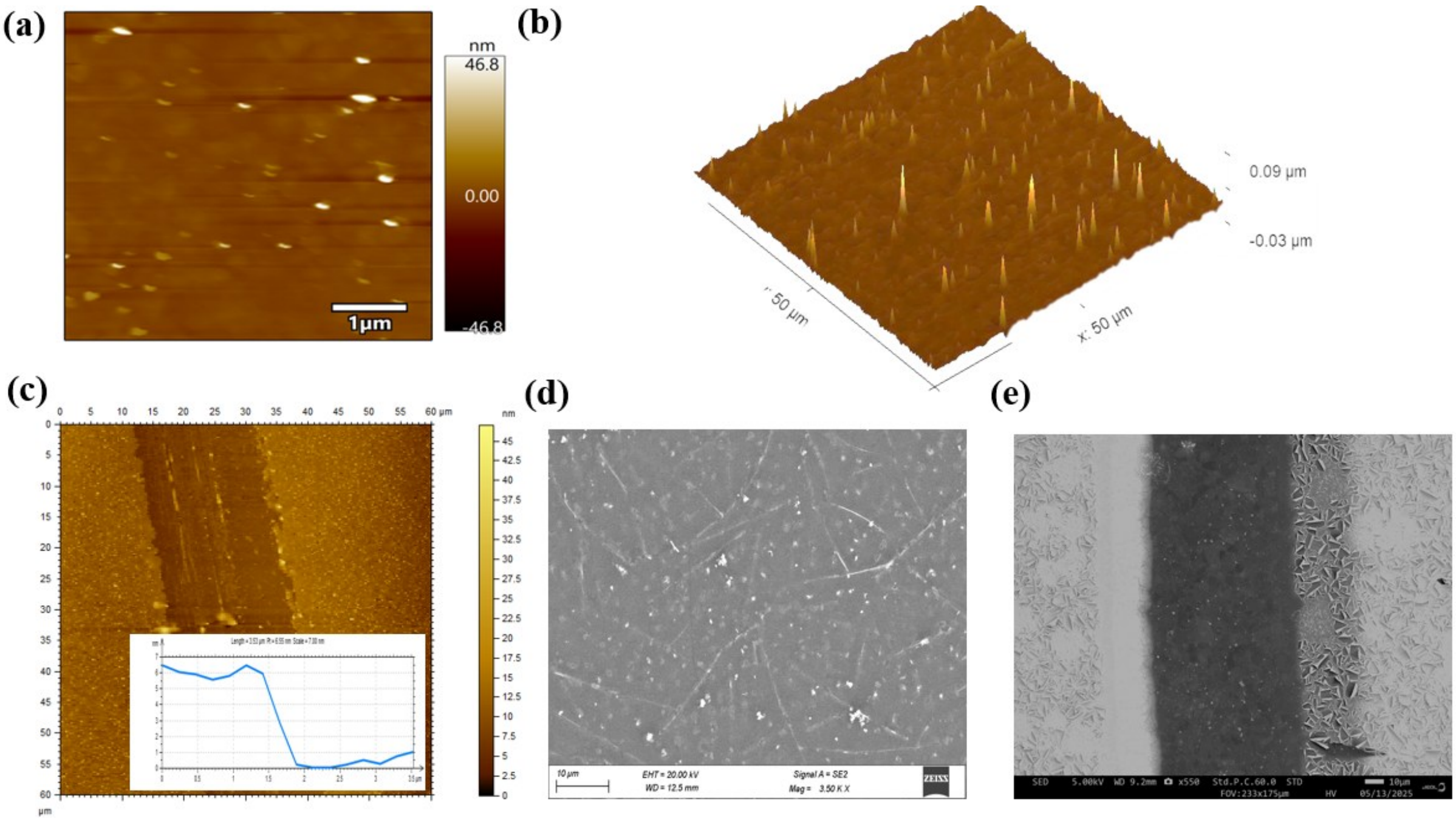

*Figure S6: (a) AFM topographic image of the as-grown film with a rms roughness of ~ 3 nm. (b) 3D image of the alloyed thin film (c) thickness of the film from AFM measurement; which is ~ 6 nm. (d) FESEM image of the film surface and (e) FESEM image of the device-channel region*

# Note 2: Stoichiometry calculation from the XPS data

The following formula has been used to estimate the stoichiometry ratio of the as-grown $MoWS_2$ alloy under ambient conditions.

$$F_x = \frac{\frac{I_x}{S_x}}{\left(\frac{I_{Mo}}{S_{Mo}}+\frac{I_W}{S_W}+\frac{I_S}{S_S}\right)} \text{.........[S4]}$$

Where $F_x$ means atomic fraction of x-element in the alloyed sample, x stands for Mo, W and S; $I_x$ and $S_x$ stand for the integrated intensity ( area under the fitted curve) and atomic sensitivity factor (ASF) of the corresponding element x[4]. These ASF values refer to the photoionization cross-section at a photon energy of 1 keV, modelled by Scofield, where $S_{Mo}$ = 3.321, $S_W$ = 3.523, and $S_S$ = 0.666. We estimated the alloy composition to be $Mo_{0.18}W_{0.16}S_{0.64}$. Thus, we have successfully grown $MoWS_2$ alloy with an intermediate composition (x = 0.5), although there is a small deviation in the concentration of the S atom ($Mo_xW_{1-x}S_2$), indicating the presence of sulfur vacancies. Here, the composition ratio (x) is 0.18. The small non-stoichiometric ratio indicates that S vacancies exist in the alloyed crystal. A large number of S atoms are exposed to the outer surface due to the thinness of the CVD layer and hence can be easily detached from the surface, consequently increasing the sulfur vacancies in the 2D CVD MoWS2 film. The presence of a high density of sulfur vacancies at extremely high growth temperatures has been previously reported[56]. Also, S has less electronegativity than oxygen, meaning that it takes less binding energy to break the Mo-S bonds compared to the Mo-O bond under ambient conditions.

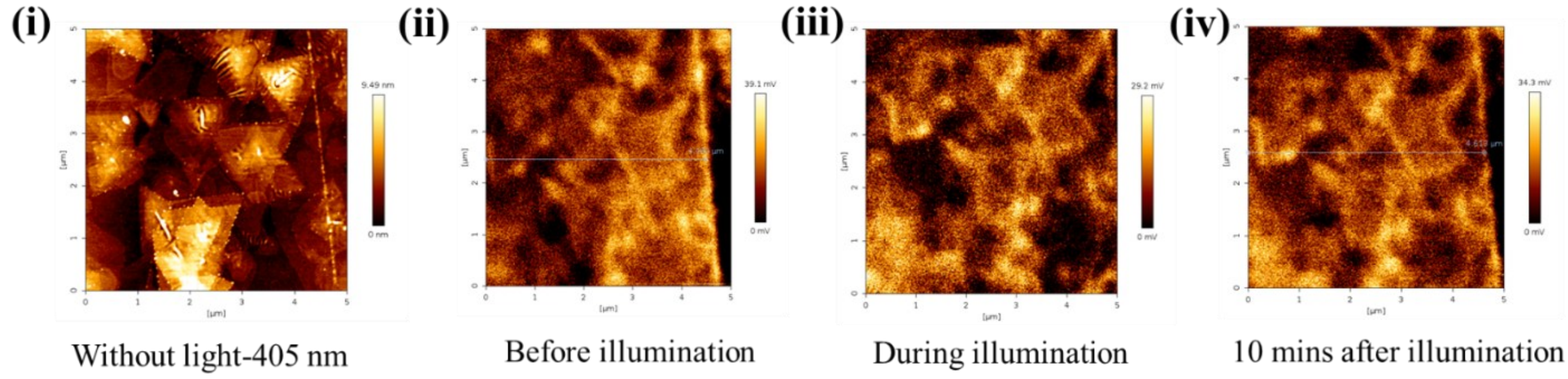

*Figure S7:(i) Height profile-AFM image of a few flakes, KFM images showing (ii) surface potential profile before light illumination (iii) during light illumination (405 nm), indicating a drastic drop in the surface potential and (iv) gradual restoration of potential after 10 mins of light removal*

## Note 3: Kelvin Probe Force Microscopy (KPFM) Analysis

Figure S7 (i) shows the typical height profile image of the as-grown $MoWS_2$ thin film. Figure S7 (ii) shows the surface potential map for the film shown in Figure S7 (i) under dark conditions, clearly indicating a remarkable variation in surface potential across the sample surface. The non-uniformity of the surface potential strongly suggests a non-uniform distribution of defect states throughout the samples, primarily due to growth-induced defect states, such as sulfur vacancies or oxygen trap centres, at the $SiO_2$/Si-$MoWS_2$ interface or at grain boundaries within the CVD film. The thicker regions with higher surface potential variation are more likely to have negatively charged entities upon light exposure. The sulfur vacancies in these thicker regions typically function as electron-trapping centres, capturing the photogenerated charge carriers upon illumination and hence resulting in a drastic reduction in surface potential. The trapped electrons are released from these trap centres over time and hence recombine with the holes. Thus, the surface potential gradually recovers to its initial level after a long relaxation time (a few minutes). The topographic image of the region and KPFM -surface potential data for the region under dark and illumination conditions are given in Figure S7.

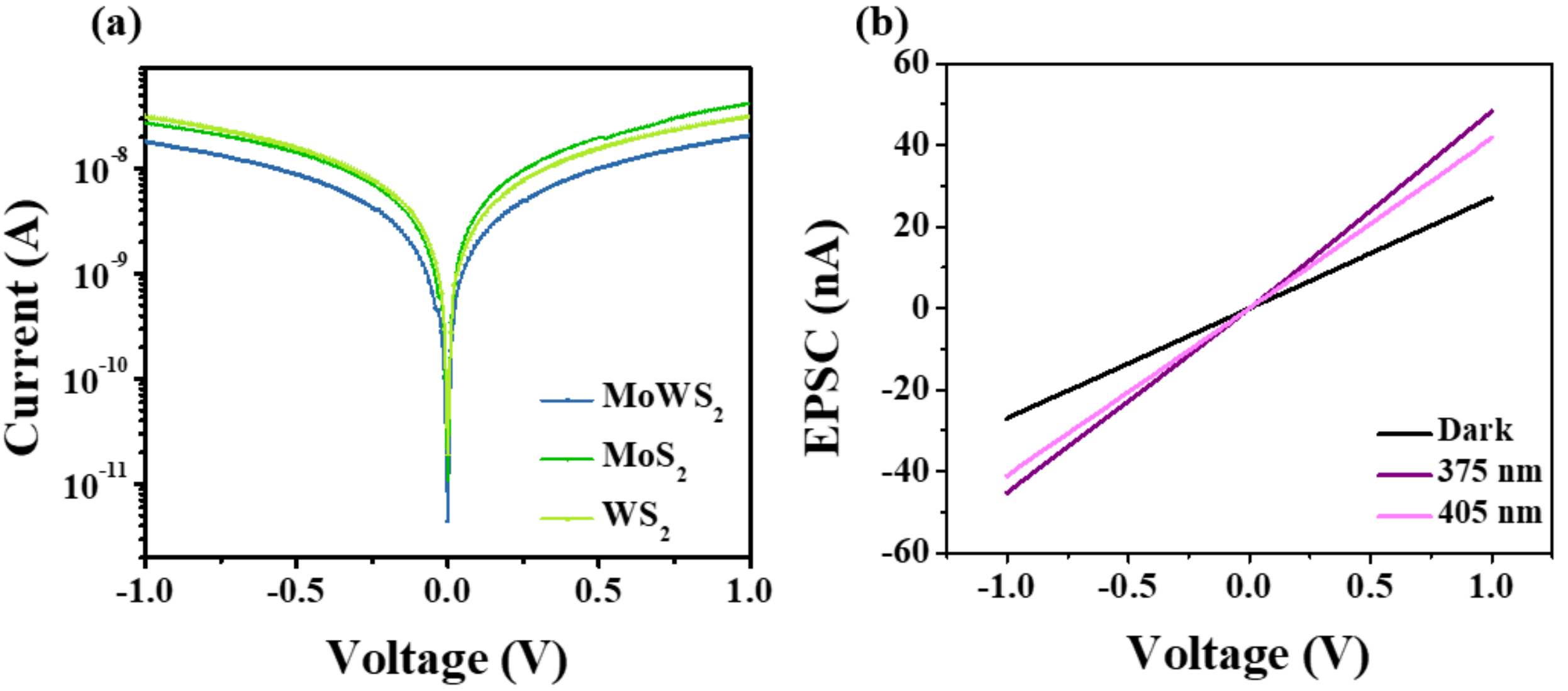

*Figure S8: (a) IV characteristics under dark conditions for $MoS_2$, $WS_2$ and $MoWS_2$ (b) Spectra IV for $MoWS_2$ OSD device under dark, 375 nm and 405 nm light.*

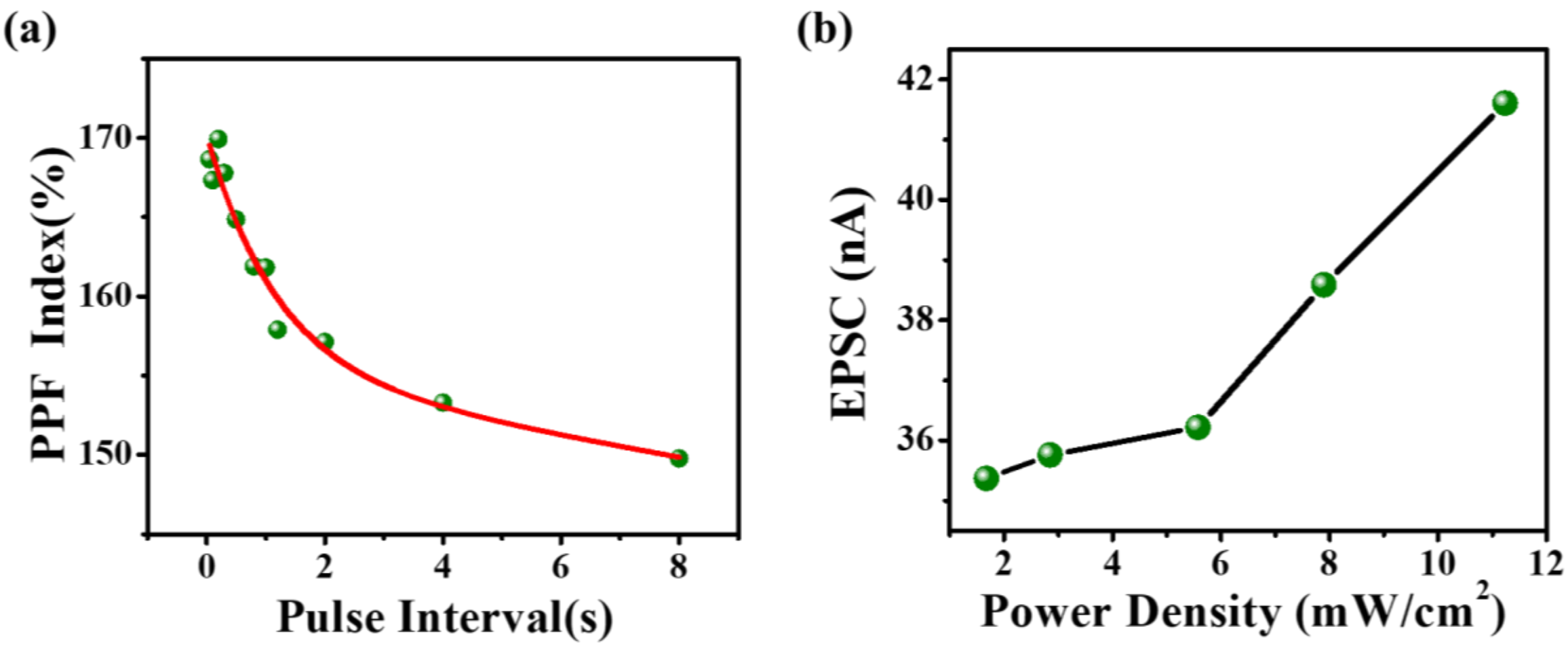

*Figure S9: (a) Dependence of PPF index on the pulse interval or pulse delay time (b) Non-linear rise of EPSC with increase of optical power density*

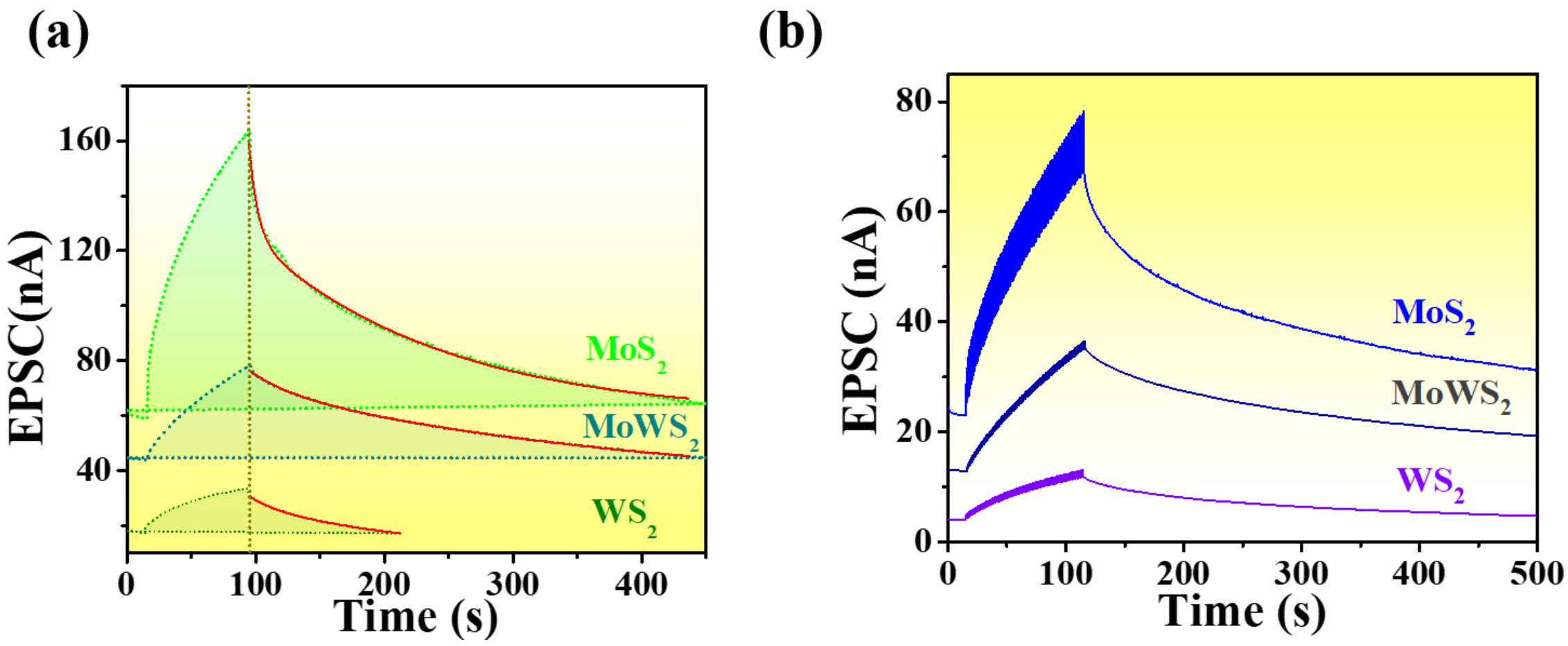

*Figure S10: (a) Optical memory retention in three different devices: $MoS_2$, $MoWS_2$ and $WS_2$ (b) Effect of 100 identical pulses on the EPSC values*

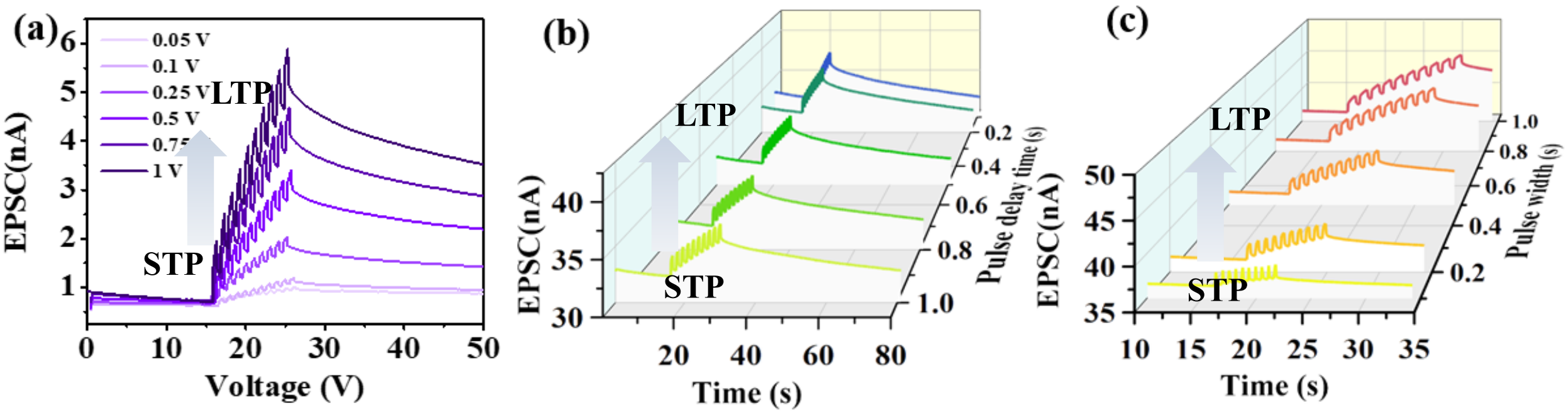

*Figure S11: STP-to-LTP transition via (a) stimulation with bias voltage (b) pulse delay time (c) Pulse width.*

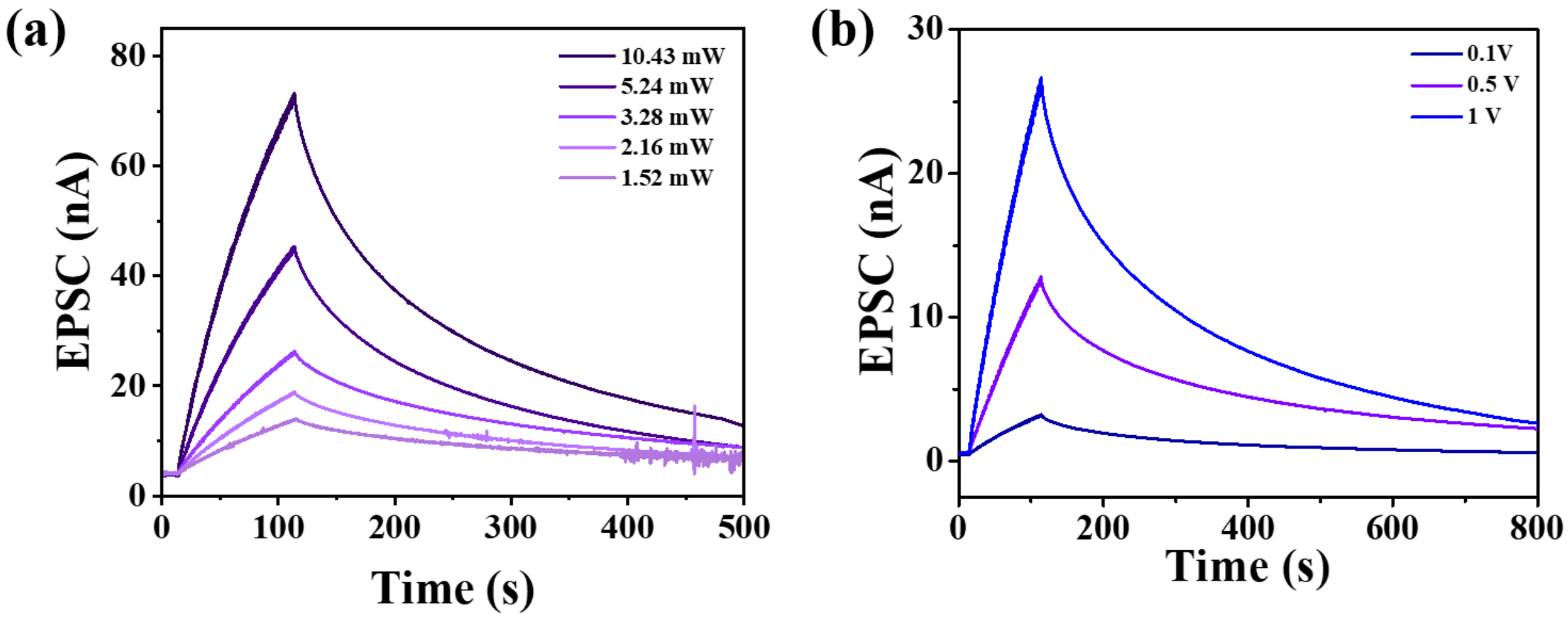

*Figure S12: (a) STP-LTP transition with increase of optical power (one optical spike of pulse width 80 s) (b) Bias-dependent STP-to-LTP transition (one optical spike of pulse width 80 s)*

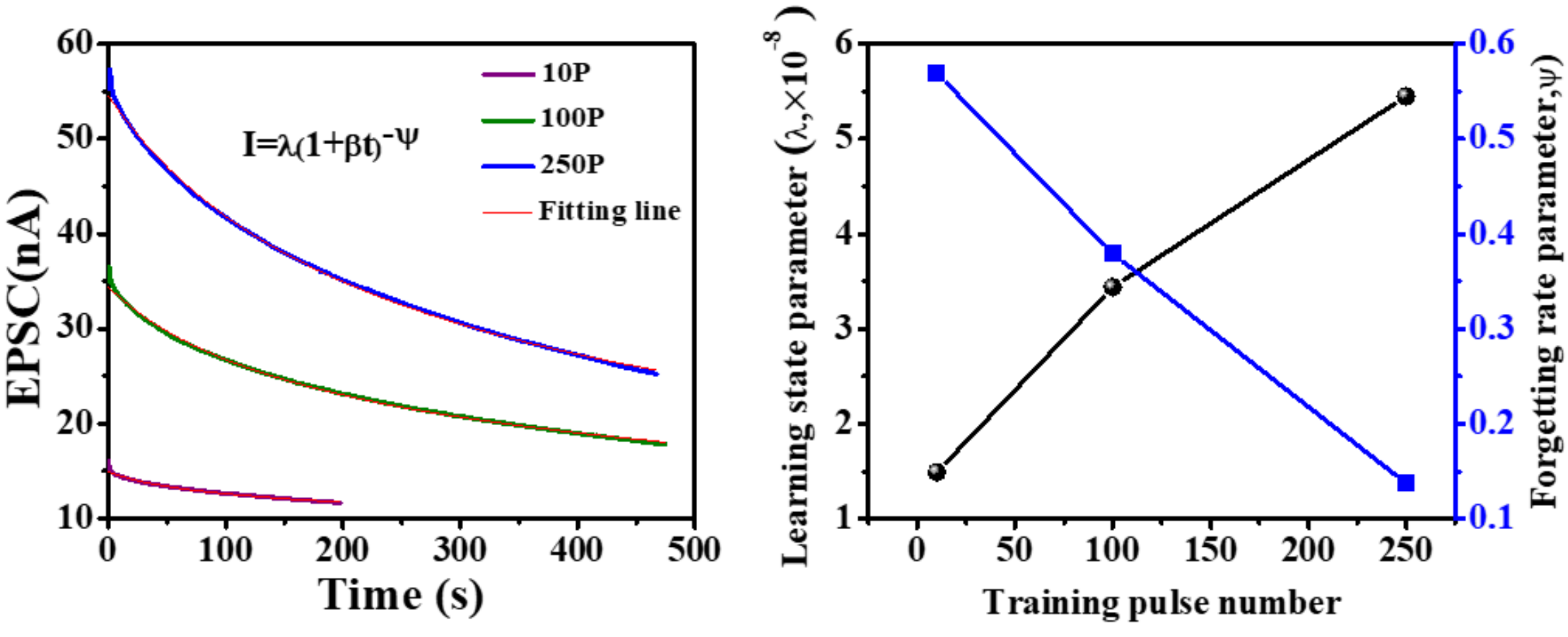

*Figure S13: (a) Fitting decay curves for 10, 100 and 250 pulses (b) Learning and forgetting parameters with varying training pulse numbers*

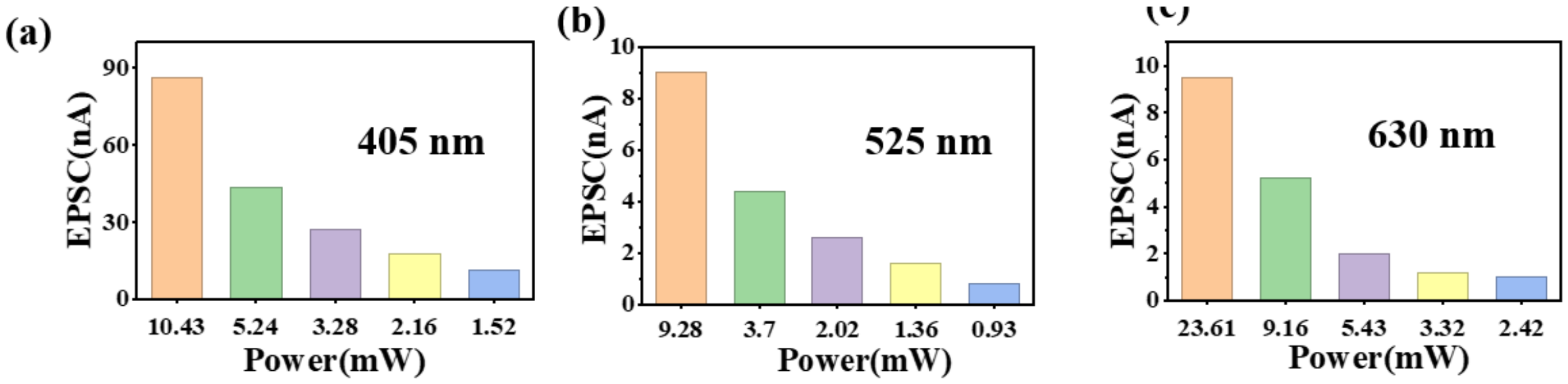

*Figure S14: Fitting of the EPSC decay curve at two different frequencies: 0.5 Hz and 2 Hz*

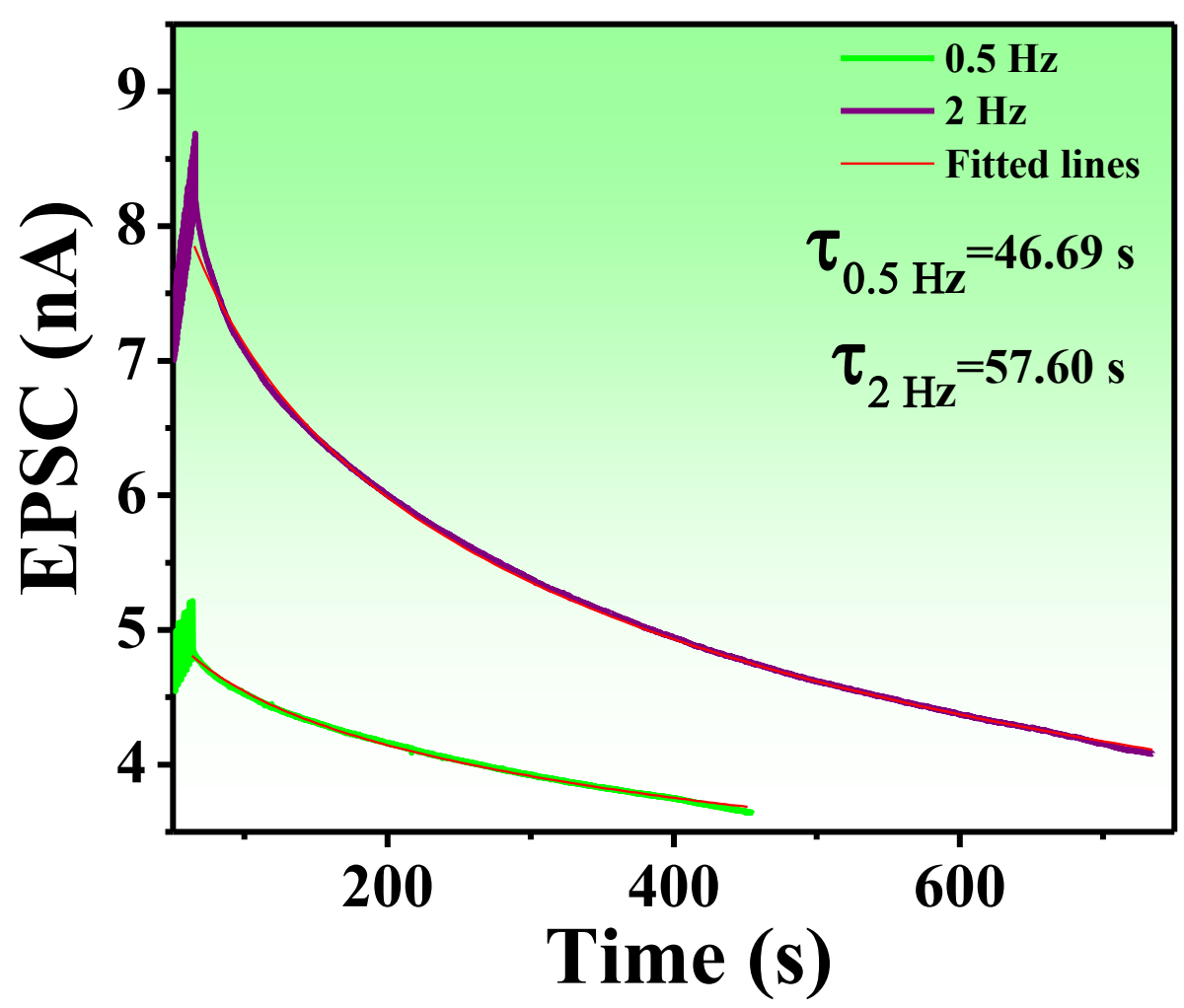

*Figure S15: Spectral photoresponse of different wavelengths at different optical powers: light stimulation on the EPSC of the device at (a) 405 nm (b) 525 nm (c) 630 nm*

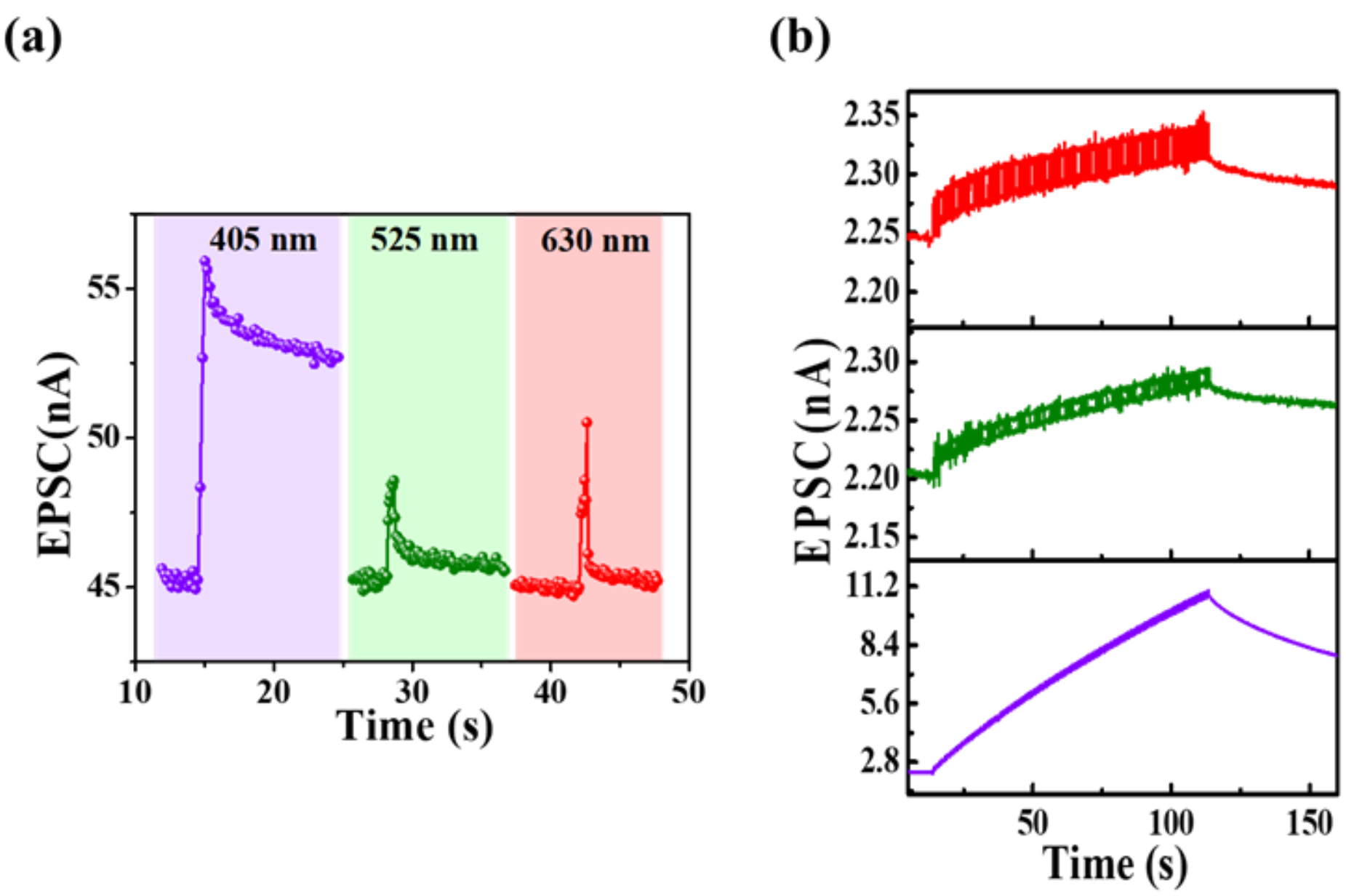

*Figure S16: (a) Dependence of the EPSC on illumination wavelength for (a) single pulse (PW=500 ms) (b) 100 pulses (PW=PD=500 ms)*

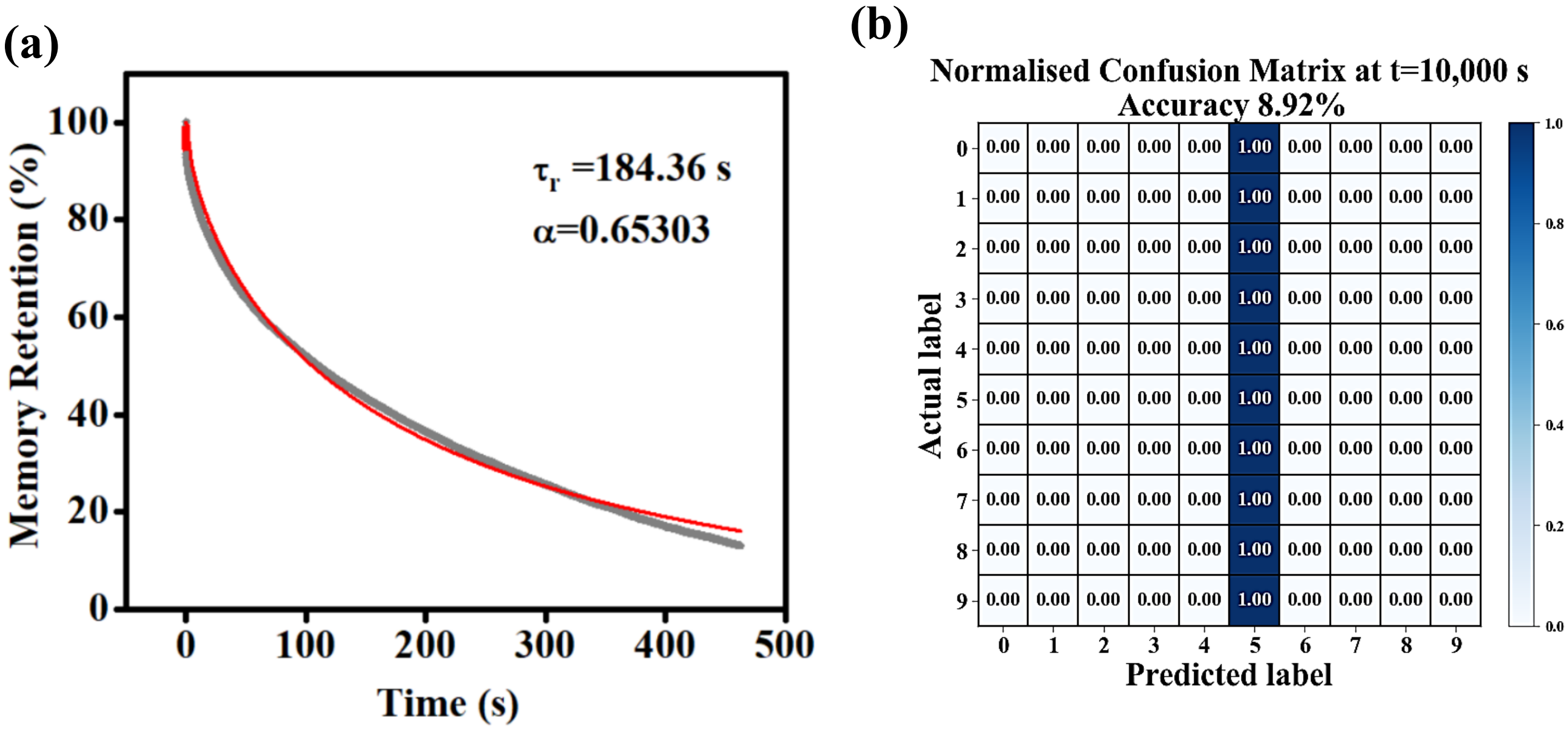

*Figure S17: (a) Memory retained (%) vs time plot; fitted with Kohl-Rausch law (b) Normalised confusion matrix at t=10,000 s with the current decayed to the base level, achieving an accuracy of 8.92% at the end of t=10,000, centred around a random digit (Here '5') depending on bias rather than the synaptic weights.*